\documentclass{jfm}
\usepackage{graphicx}
\usepackage{epstopdf, epsfig}

\shorttitle{Baroclinic acoustic streaming}
\shortauthor{G. Michel and G. P. Chini}

\title{Strong wave--mean-flow coupling in baroclinic acoustic streaming}

\author{Guillaume Michel\aff{1}
  \and Gregory P. Chini\aff{2}
  \corresp{\email{greg.chini@unh.edu}}}

\affiliation{%\aff{1}Woods Hole Oceanographic Institution, Woods Hole, MA 02543, USA
\aff{1} Laboratoire de Physique Statistique, \'Ecole Normale Sup\'erieure, CNRS, Universit\'e P. et M. Curie, Universit\'e Paris Diderot, Paris 75005, France
\aff{2}Department of Mechanical Engineering and Program in Integrated Applied Mathematics, University of New Hampshire, Durham, NH 03824, USA}

\begin{document}

\maketitle

\begin{abstract}
The interaction of an acoustic wave with a stratified fluid can drive strong streaming flows owing to the baroclinic production of fluctuating vorticity, as recently demonstrated by Chini \emph{et al.} (\emph{J. Fluid Mech.}, \textbf{744}, 2014, pp. 329--351).  In the present investigation, a set of wave/mean-flow interaction equations is derived that governs the coupled dynamics of a standing acoustic-wave mode of characteristic (small) amplitude $\epsilon$ and the streaming flow it drives in a thin channel with walls maintained at differing temperatures.  Unlike classical Rayleigh streaming, the resulting mean flow arises at $\mathit{O}(\epsilon)$ rather than at $\mathit{O}(\epsilon^2)$.  Consequently, fully two-way coupling between the waves and the mean flow is possible:  the streaming is sufficiently strong to induce $\mathit{O}(1)$ rearrangements of the imposed background temperature and density fields, which modifies the spatial structure and frequency of the acoustic mode on the streaming time scale.  A novel Wentzel--Kramers--Brillouin--Jeffreys analysis is developed to average over the fast wave dynamics, enabling the coupled system to be integrated strictly on the slow time scale of the streaming flow.  Analytical solutions of the reduced system are derived for weak wave forcing and are shown to reproduce results from prior direct numerical simulations (DNS) of the compressible Navier--Stokes and heat equations with remarkable accuracy.  Moreover, numerical simulations of the reduced system are performed in the regime of strong wave/mean-flow coupling for a fraction of the computational cost of the corresponding DNS.  These simulations shed light on the potential for baroclinic acoustic streaming to be used as an effective means to enhance heat transfer.
\end{abstract}


\section{Introduction}
Sound waves can drive Eulerian flows that evolve on a slow time scale compared to the period of the waves. The theoretical study of this phenomenon, called acoustic (or, in other contexts, steady) streaming, can be traced back to Rayleigh in the 19th century \citep{Rayleigh1884}.  Given that ultrasonic power sources are now routinely used in laboratory experiments, acoustic streaming has been widely observed, often as a source of unwanted flow. Nonetheless, streaming also has been recognized as a practical means to enhance transport and mixing and has, for instance, been used to improve the efficiency of chemical reactions occurring near a catalytic solid phase that otherwise would be controlled by molecular diffusion \citep{Bengtsson2004}; to directly mix chemical species \citep{Yaralioglu2004}; and for activated irrigation in medical applications including root-canal procedures \citep{Verhaagen2014}.  Heat also can be transported by streaming flows, and acoustic waves therefore can be used to accelerate the cooling of a hot object, as recently reviewed by \cite{Legay2011}.  Acoustic streaming technologies are of particular interest in the zero-gravity environment, where natural convective flows do not exist and acoustic thermal management systems may provide a reliable, efficient and lightweight alternative to fans. 

%In a characteristically lucid lecture, Lighthill identified the different regimes of acoustic streaming in a homogeneous medium \citep{Lighthill1978}.  Owing to  attenuation mechanisms, acoustic waves generate a Reynolds stress divergence capable of driving a mean flow, which is balanced either by viscosity (e.g. acting in oscillatory boundary layers, so-called ``Rayleigh streaming'') or by inertia (``Stuart streaming''). 

In a characteristically lucid lecture, Lighthill identified the different regimes of acoustic streaming in a homogeneous medium \citep{Lighthill1978}.  Owing to  attenuation mechanisms, acoustic waves generate a Reynolds stress divergence capable of driving a mean flow, which is balanced either by viscous forces (termed ``Rayleigh streaming'' if, in addition, the sound waves are damped in oscillatory boundary layers) or by inertia (``Stuart streaming''). The former regime occurs for small values of the streaming Reynolds number $Re_s = U_sL/\nu$, where $U_s$ is a characteristic streaming speed, $L$ is a typical dimension of the system and $\nu$ is the kinematic viscosity of the fluid. 
%If the acoustic waves are damped in oscillatory boundary layers, this Rayleigh streaming comprises a laminar flow localized within a few wavelengths of any solid boundary  \citep{Nyborg1958} and can be analytically computed for simple geometries, e.g. in a channel \citep{Rayleigh1884, Hamilton2003} or adjacent to a circular cylinder \citep{Holtsmark1954}.  
In Rayleigh streaming, the (laminar) cellular mean flow that is generated is localized within a few wavelengths of any solid boundary \citep{Nyborg1958}; the streaming can be analytically computed for simple geometries, e.g. in a channel \citep{Rayleigh1884, Hamilton2003} or adjacent to a circular cylinder \citep{Holtsmark1954}. This regime has become important in microfluidics, where the vortices induced by acoustic streaming in microchannels can be used to mix chemicals (see references above).  In contrast, for large $Re_s$, the streaming flow acquires a jet-like structure and can become turbulent \citep{Stuart1966,Lighthill1978}.

The presence of an inhomogeneous background temperature (or density) field strongly affects the fundamental mechanics and kinematics of acoustic streaming:  streaming velocities are significantly enhanced and the flow patterns are substantially altered.  These changes are evident in the early experiments of \cite{Fand1960} and in subsequent experiments and numerical simulations; see e.g. \cite{Loh2002}, \cite{Hyun2005}, \cite{Lin2008}, \cite{Nabavi2008}, \cite{Aktas2010}, \cite{Dreeben2011} and \cite{Karlsen2017}.  Consequently, the resulting flow and associated transport cannot be computed simply by coupling the corresponding isothermal (e.g. Rayleigh or Stuart) streaming with the heat or other appropriate transport equation.  Instead, new physical phenomena occur, which renders this problem both complex and interesting.  Experimental challenges arise because natural convection and acoustic streaming may be difficult to disentangle in the laboratory; numerical challenges result from the need to resolve compressible fluid dynamics on temporal scales ranging from the acoustic wave period to the slow time scale over which the streaming flow evolves; while the primary theoretical challenge is to elucidate the novel phenomenology resulting from fully two-way coupling between the sound waves and the mean flow.

This striking change in the character of the streaming has been observed in various contexts in which an agency \emph{other} than viscosity generates vorticity in the oscillatory flow (e.g. the acoustic waves). \cite{Amin1988} and \cite{Riley_Trinh2001} noted fundamental changes in the steady streaming driven by a non-conservative body force in their study of fluid flow in the presence of $g$-jitter, i.e. a fluctuating gravitational field in an otherwise gravity-free environment.   Motivated by the observation that streaming velocities in high-intensity discharge lamps are two orders of magnitude larger than those predicted by Rayleigh streaming theory \citep{Dreeben2011}, \cite{Chini2014} derived a theory capable of accounting for both the observed streaming pattern and magnitude.  As discussed more fully in \S~\ref{S1}, the mechanism underlying this large-amplitude streaming is the baroclinic production of sound-wave vorticity arising from the misalignment of fluctuating isobars and mean isopycnals.  Similarly, non-classical streaming phenomena have been observed in microfluidic systems with gradients in density; in particular, \cite{Karlsen2016, Karlsen2017} recently obtained a local expression for the acoustic force density driving the streaming flow as a function of the acoustic-wave characteristics.

The primary objective of the present investigation is to systematically extend the recent theory of \cite{Chini2014} to efficiently capture the two-way coupling that can occur in baroclinic acoustic streaming and to quantify the concomitant heat transfer.  Accordingly, a novel Wentzel--Kramers--Brillouin--Jeffreys (WKBJ) analysis is performed to enable prediction of the slow evolution of the acoustic wave amplitude, thereby obviating the need to explicitly simulate the fast oscillatory dynamics.  We focus on perhaps the most well-documented acoustic streaming configuration:  a thin two-dimensional channel with an imposed standing acoustic wave oscillating in the wall-parallel direction.  For a homogeneous system, the resulting streaming flow was first described in the pioneering work of \cite{Rayleigh1884} in the limits $Re_s\ll 1$ and $\delta_* \ll H_* \ll k_*^{-1}$, where $\delta_* = \sqrt{2\nu_* / \omega_*}$ is the thickness of the oscillatory (Stokes) boundary layers ($\nu_*$ is the kinematic viscosity, and $\omega_*$ is the wave angular frequency), $H_*$ is the channel width and $k_*$ is the wavenumber of the acoustic wave. The streaming flow comprises a wall-parallel array of counter-rotating vortices, stacked in the wall-normal direction and having a characteristic velocity $3U_*^2 / (16a_*)$, where $U_*$ is the maximum fluctuating velocity induced by the standing acoustic wave and $a_*$ is the speed of sound. Experiments were first performed in a tube and showed quantitative agreement with the predictions of Rayleigh \citep{Andrade1931}. \cite{Hamilton2003} extended this theoretical study to channels of arbitrary width $H_*$, the only restrictions being $Re_s \ll 1 $ and $\delta_* \ll  k_*^{-1}$.  When the upper and lower walls of the channel are maintained at fixed but differing temperatures, both experiments and direct numerical simulations of the compressible Navier--Stokes and heat equations indicate a change in the streaming phenomenology:  the stacked vortices merge and their characteristic velocity increases \citep{Loh2002, Lin2008}.  To date, no theory has correctly predicted the resulting streaming-flow pattern and intensity; our study, which focuses on the regime $Re_s \gtrsim 1$, fills this gap in the literature.

The remainder of the paper is organized as follows. After formulating the problem for the instantaneous dynamics, we carry out a multiple scale analysis (\S~\ref{S1}) to obtain a reduced but two time-scale system.  In \S~\ref{S2}, we analyse the wave dynamics to show that this multiscale system can be integrated strictly on the slow time scale. We then consider, in \S~\ref{S3}, the limit of weak wave forcing, in which the streaming flow does not produce appreciable feedback on the waves; in particular, we derive an approximate analytical solution and compare it to the streaming flow numerically computed by \cite{Lin2008}, demonstrating excellent quantitative agreement.  In \S~\ref{S4}, we perform numerical simulations of our reduced model and characterize the resulting fully-coupled waves and mean flows. We summarize our key findings and suggest possible further extensions in~\S~\ref{S5}.

%\section{Multiple time scale formulation\label{S1}}
\section{Two time-scale wave/mean-flow system\label{S1}}

\subsection{Flow configuration}

\begin{figure}
  \centerline{\includegraphics{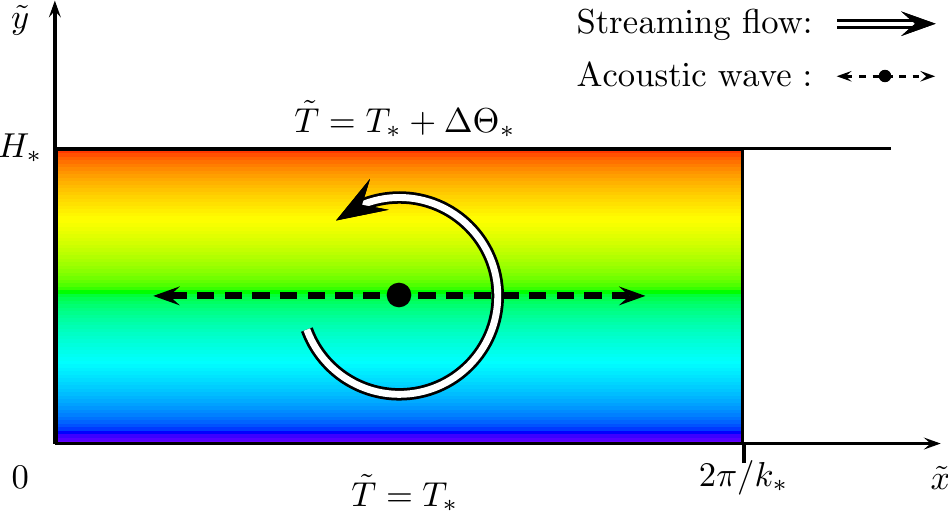}}
  \caption{Schematic of the flow configuration.  A thermally-stratified ideal gas is confined between the plane parallel walls of a long thin channel of height $H_*$.  A standing acoustic wave of wavelength $2\pi/k_*$ interacts with the thermal stratification to drive a time-mean, or streaming, flow that is sufficiently strong to modify the wave dynamics.  The thermal driving is imposed by fixing the temperature $\tilde{T}$ of the lower wall to be $T_*$ and that of the upper wall to be $T_*+\Delta\Theta_*$.}
\label{fig:schematic}
\end{figure}

The problem we consider is similar to that introduced in \cite{Chini2014}.  Specifically, we analyze the two-dimensional flow of an ideal gas, with specific gas constant $R_*$ and constant dynamic viscosity $\mu_*$ and thermal conductivity $\kappa_*$, in a channel with walls separated by a distance $H_*$ in the $\tilde{y}$ coordinate direction (see figure~\ref{fig:schematic}).  Here and throughout, tildes refer to dimensional variables, while asterisks are used to denote dimensional parameters.  Subsequently, overbars will be used to designate dimensionless time-averaged fields, while primes will be reserved for dimensionless oscillatory fields.  The gas is presumed to be driven in an approximately time-periodic fashion, with frequency $\omega_*$, yielding a standing sound wave with spatial wavenumber $k_*$.  The velocity field is required to satisfy no-slip  and zero normal-flow boundary conditions along the channel walls located at $\tilde{y}=0$ and $\tilde{y}=H_*$.  All dependent fields are required to satisfy a $2\pi/k_*$ periodicity condition in the horizontal ($\tilde{x}$) coordinate.  In addition, to fix the spatial phase of the sound wave, we impose a symmetry condition along (or, equivalently, a zero mass-exchange condition across) $\tilde{x}=0$; i.e. $\tilde{u}(0,\tilde{y},\tilde{t})$=0, where $\tilde{u}$ is the $\tilde{x}$-velocity component and $\tilde{t}$ is the time variable.  This additional boundary condition holds for any flow developing from initial conditions that are symmetric with respect to $\tilde{x}=0$, e.g. for the quiescent diffusive state, since then both the initial conditions and the governing equations are invariant with respect to the transformation $\tilde{x} \rightarrow -\tilde{x}$, $\tilde{u} \rightarrow -\tilde{u}$ and $\tilde{v} \rightarrow \tilde{v}$.

In contrast to the study of \cite{Chini2014}, the thermal driving is achieved by fixing the temperatures of the lower and upper walls to be $T_*$ and $T_* + \Delta \Theta_*$, respectively, rather than by including a volumetric heat source.  For convenience, we take the temperature differential $\Delta \Theta_* > 0$, but note that this restriction is not dynamically significant since we do not consider the influence of buoyancy (gravity) in this investigation.  Denoting the density, pressure, temperature and velocity fields by $\tilde{\rho}$, $\tilde{p}$, $\tilde{T}$ and $\tilde{\mathbf{u}}$, respectively, where $\tilde{\mathbf{u}}=(\tilde{u},\tilde{v})$ and $\tilde{v}$ is the $\tilde{y}$-velocity component, the governing (compressible) Navier--Stokes, continuity and energy equations and the ideal gas equation of state can be written as
%It consists of a two-dimensional channel sketched in figure \ref{fig:schematic} of thickness $H_*$ along the vertical $\tilde{y}$ direction and of periodicity $2\pi / k_*$ along the horizontal $\tilde{x}$ direction. In all the following, tildes and stars refer to dimensional quantities, bars to time-averaged quantities and primes to oscillating fields. Table \ref{tab:dimensional_parameters} summarizes all the dimensional parameters introduced. The fluid is assumed to be an ideal gas of specific gas constant $R_s$, whose dynamic viscosity $\mu$ and thermal conductivity $\kappa$ are independent of the temperature. The kinematic boundary conditions are no slip boundary conditions at $\tilde{y}=0$ and $\tilde{y}= H_*$. Moreover, we require that there is no exchange of mass between nearby cells and fix $\tilde{u}(\tilde{x} = 0, \tilde{y}, \tilde{t}) = 0$, where $\tilde{u}$ is the horizontal velocity. The thermal boundary conditions consist in a constant temperature at the bottom, $\tilde{T}(\tilde{x}, \tilde{y}=0, \tilde{t}) = T_*$, and at the top, $\tilde{T}(\tilde{x}, \tilde{y}=H_*, \tilde{t}) = T_* + \Delta \Theta_*$, where $\Delta \Theta_* > 0$. Finally, we do not consider any gravity field, so that the restriction $\Delta \Theta_* > 0$ is only a matter of clarity.
\begin{equation} \label{D_momentum}
\tilde{\rho} \left[ \partial_{\tilde{t}} \tilde{\mathbf{u}} + ( \tilde{\mathbf{u}} \cdot \tilde{\mathbf{\nabla}} )\tilde{\mathbf{u}} \right] = - \tilde{\mathbf{\nabla}} \tilde{p}  +  \mu_* \left[ \tilde{\mathbf{\nabla}}^2 \tilde{\mathbf{u}} + \frac{1}{3}\tilde{\mathbf{\nabla}}  ( \tilde{\mathbf{\nabla}} \cdot \tilde{\mathbf{u}} ) \right],
\end{equation}
\begin{equation} \label{D_mass}
\partial_{\tilde{t}} \tilde{\rho} + \tilde{\mathbf{\nabla}}  \cdot \left( \tilde{\rho} \tilde{\mathbf{u}}  \right) = 0,
\end{equation}
\begin{equation} \label{D_heat}
\tilde{\rho} c_v  \left[ \partial_{\tilde{t}} \tilde{T} + ( \tilde{\mathbf{u}} \cdot \tilde{\mathbf{\nabla}} )\tilde{T} \right] = - \tilde{p} \left( \tilde{\mathbf{\nabla}} \cdot \tilde{u} \right)  + \kappa_* \tilde{\mathbf{\nabla}}^2 \tilde{T},
\end{equation}
\begin{equation} \label{D_state}
\tilde{p} = \tilde{\rho} R_* \tilde{T},
\end{equation}
where the two-dimensional gradient operator $\tilde{\nabla} = (\partial_{\tilde{x}}, \partial_{\tilde{y}})$.  Note that dilatational (or `bulk') viscosity has been neglected in (\ref{D_momentum}), and viscous heating has been omitted in (\ref{D_heat}).  In practice, the bulk viscosity vanishes for a monatomic gas and, according to early experiments, is smaller than the dynamic (or shear) viscosity for the specific diatomic ideal gas (i.e. nitrogen) studied here \citep{Prangsma1973}. More importantly, although significant variations in the dynamic viscosity may be expected owing to the temperature dependence of this coefficient, these variations are neglected to facilitate the analysis.
	\begin{table}
		\begin{center}
		\def~{\hphantom{0}}
    \begin{tabular}{cc}
	       \underline{Notation} & \underline{Definition}\\
	    $\tilde{\mathbf{u}}= (\tilde{u},\tilde{v})$ & Gas velocity\\	       
	    $\tilde{\rho}$ & Gas density\\
	    $\tilde{p}$ & Gas pressure\\
	    $ \tilde{T}$ & Gas temperature\\
	    $(\tilde{x},\tilde{y})$ & Horizontal, vertical (wall-normal) coordinate\\
	    $\tilde{t}$ & Time variable\\
	    $H_*$ & Channel height\\
	    $k_*$ & Acoustic-wave horizontal wavenumber\\
	    $\mu_*$ & Dynamic viscosity\\
	    $\kappa_*$ & Thermal conductivity\\
	    $R_*$ & Specific gas constant\\
	    $(c_{v_*}, c_{p_*})$ & Constant volume, pressure specific heat coefficient\\
	    $a_*=\sqrt{(c_{p_*}/c_{v_*}) R_* T_*}$ & Background sound speed\\
	    $p_*$ & Background pressure field\\
%	    $U_*$ & Typical velocity amplitude of the acoustic wave\\
		\end{tabular}    
	\end{center}
	\caption{Dimensional variables and parameters}
%		\caption{Definitions of the dimensional fields and parameters}
	\label{tab:dimensional_parameters}
\end{table}
%
%The dimensional equations are reported below, with $\tilde{\nabla} = (\partial_{\tilde{x}}, \partial_{\tilde{y}})$ : 
%
%
%
%
%Equation (\ref{D_momentum}) is the Navier-Stokes equation in which volume viscosity has been neglected. Equation (\ref{D_mass}) stands for the conservation of mass, (\ref{D_heat}) is the heat equation (viscous heating is neglected) and (\ref{D_state}) is the equation of state. 
The steady-state pressure and temperature fields in the absence of acoustic waves and streaming flow, that is for $\tilde{\mathbf{u}}= \mathbf{0}$, are referred to as the background fields (denoted with a subscript `$B$') and are found to be
\begin{equation}
\tilde{T}_B = T_* \left( 1 + \Gamma \frac{y_*}{H_*} \right),~~~~~ \tilde{p}_B = p_*, \label{Background_fields}
\end{equation}
where $p_*$ is a constant and the dimensionless temperature differential $\Gamma\equiv\Delta\Theta_*/T_*$.
%is defined in the following subsection.  
Table~\ref{tab:dimensional_parameters} summarizes the dimensional fields and parameters used in the following analysis.

\subsection{Scaling and non-dimensionalisation}
To facilitate the asymptotic analysis, we non-dimensionalise the governing equations by scaling the dependent and independent variables as outlined in table~\ref{tab:scalings}. The $\tilde{x}$-velocity component is scaled with $a_*$, the sound speed at temperature $T_*$. This scaling introduces into the dimensionless governing equations the Strouhal number $S = a_* / U_*$, where $U_*$ (rather than $a_*$) is a characteristic oscillatory velocity induced by the standing acoustic wave.  The Strouhal number is large ($10^3$ or larger) in many applications, and therefore we introduce $\epsilon \equiv 1/ S$ and consider the asymptotic limit $\epsilon\to 0$ with all other dimensionless parameters scaled as appropriate powers of $\epsilon$.  Since $\epsilon\ll 1$, the leading-order acoustic wave dynamics is linear.  Nevertheless, weak wave--wave nonlinearities are crucial for acoustic streaming, as their cumulative effect can be significant over sufficiently many [$\mathit{O}(1/\epsilon)$] acoustic wave periods (i.e. over the slow time scale).  The implied temporal scale separation between the wave and streaming dynamics is readily achieved in both laboratory experiments and streaming-enabled technologies.

The $\tilde{x}$ and $\tilde{y}$ coordinates are scaled with $k_*^{-1}$ and $H_*$, respectively, so that the gas lies in the domain defined by $x\in [0,2\pi ]$ and $y \in [0,1]$.  Time $\tilde{t}$ is non-dimensionalised using the inverse reference wave frequency $\omega_*^{-1}=(a_*k_*)^{-1}$.  The vertical ($\tilde{y}$), or wall-normal, velocity component is scaled by $(k_*H_*)a_*$.  The domain aspect ratio $\delta\equiv k_* H_*$ is assumed to be small and, more precisely, is chosen so that $\delta = \sqrt{\epsilon} h$, where $h$ is a dimensionless parameter of order unity.    Although the cross-channel heat flux is not expected to be maximized as $\delta\to 0$ (since the streaming flow will be largely horizontal), we follow \cite{Chini2014} and continue to focus on the small aspect-ratio regime for the following reasons.  First, as noted in the introduction, most theoretical and computational studies have been performed in this regime, so meaningful comparisons to prior investigations can be made. Secondly, the analysis of the acoustic wave is simplified in a domain that is thin relative to the wavelength of the sound wave.  Indeed, the acoustic wave then is dynamically constrained to maintain its first-mode wall-normal structure.  Finally, in the small aspect-ratio regime, the leading-order fluctuating pressure gradient is orthogonal to the imposed background density gradient, resulting in a crucial baroclinic contribution to the production of fluctuating vorticity.

	\begin{table}
		\begin{center}
		\def~{\hphantom{0}}
    \begin{tabular}{c c | c  c c}
	       \underline{Variable} &  \underline{Scale} &  \underline{Parameter} &  \underline{Definition} &  \underline{Scaling}\\ 
	    $x$ & $k_*^{-1}$ & Strouhal number $S$ & $a_* / U_*$  &$S = 1/\epsilon$\\
	    $y$ & $H_*$& Aspect ratio $\delta$ & $k_* H_*$ &$\delta = \sqrt{\epsilon} h$\\
	    $t$ & $(a_* k_*)^{-1}$ & Temperature gradient $\Gamma$ & $\Delta \Theta_* / T_*$  & $\Gamma = O(1)$\\
	    $u$ & $a_*$ & Reynolds number $Re$ & $\rho_* U_*/ (k_* \mu_*)$ & $Re = Re_s /\epsilon$\\
	    $v$ & $(k_*H_*)a_*$ & P\'eclet number $Pe$ & $\rho_* c_{p_*} U_* / (k_* \kappa_*)$  & $Pe = Pe_s /\epsilon$\\
	    $\rho$ & $\rho_* \equiv p_* /(R_* T_*)$ & Specific heat ratio $\gamma$ & $c_{p_*}/c_{v_*}$ & $\gamma = O(1)$\\
	    $T$ & $T_*$&&& \\
	    $P$ & $p_*$& & & \\
		\end{tabular}    
	\end{center}% \vspace{-.5cm}
	\caption{Dimensionless variables and parameters}
%		\caption{Definitions and scalings of the dimensionless variables and parameters}
	\label{tab:scalings}
\end{table}

The temperature $T_*$ of the lower wall is used to nondimensionalise the temperature field $\tilde{T}$.  
%The thermal driving is therefore characterized by the dimensionless parameter $\Gamma\equiv \Delta \Theta_* / T_*$.  
In the analysis that follows, $\Gamma$ is fixed, i.e. $\mathit{O}(1)$, as $\epsilon\to 0$, although the smallness of $\epsilon$ in acoustic streaming ensures that our multiple scale analysis remains accurate even for $\Gamma\approx 0.1$, as will be evident in \S~\ref{S3}, where we compare our theoretical predictions with the results of direct numerical simulations.
%Such a temperature difference is large but not irrelevant for the cooling of electronics in space application. 

The Reynolds and P\'eclet numbers characterizing the acoustic waves are denoted $Re$ and $Pe$, respectively.  Since these parameters are very large compared to unity, both momentum and thermal diffusion can be neglected in the leading-order wave dynamics, at least in the domain interior. Note that in baroclinic acoustic streaming, typical streaming velocities also are of size $U_*$ (in contrast to Rayleigh streaming, where streaming speeds are proportional to $\epsilon U_*$).  Consequently, $Re$ and $Pe$ also would appear to characterize the streaming flow.  Because we consider the limit of small aspect ratio, however,  diffusion in the wall-normal ($y$) direction is enhanced by a factor $\delta^{-2} \propto \epsilon^{-1}$. We therefore define the streaming Reynolds and P\'eclet numbers $Re_s = \epsilon Re$ and $Pe_s = \epsilon Pe$ as in traditional Rayleigh streaming as the proper measure of mean inertia to the dominant mean diffusive effects.

\subsection{Asymptotic analysis}
Using the scalings described above (and summarized in table~\ref{tab:scalings}), the governing equations and boundary conditions can be recast in dimensionless form.  The occurrence of the small parameter $\epsilon$ in the dimensionless system prompts a multiple scale asymptotic analysis in which the single time variable $t$ characterizing the fast dynamics of the acoustic waves is augmented with a slow time variable $T\equiv\epsilon t$ to capture the cumulative effect of weakly nonlinear wave dynamics that ultimately drives streaming.  Furthermore, we posit the following asymptotic expansions for the various fields:
\begin{eqnarray}
(u,v)&=&\epsilon (u_1,v_1) + \epsilon^2 (u_2,v_2) + O(\epsilon^3), \label{scaling:u}\\
\pi&=&\epsilon \pi_1 + \epsilon^2 \pi_2 + O(\epsilon^3),\\
\Theta&=&\Theta_0 + \epsilon \Theta_1 + O (\epsilon^2),\\
\rho&=&\rho_0 + \epsilon \rho_1 + O (\epsilon^2),
\end{eqnarray}
where $\pi = (\tilde{p}- \tilde{p}_B) / p_*$ and $\Theta = (\tilde{T} - \tilde{T}_B) / T_*$ are, respectively, the dimensionless perturbation pressure and temperature fields. Note that, at each order in these expansions, the field variables can have both fluctuating and mean components, which subsequently will be disentangled via the introduction of a fast-time averaging operation.  Expansion (\ref{scaling:u}) follows from the scaling of the Strouhal number ($S = 1 / \epsilon$).  The state equation constrains the temperature and density perturbations to be of the same order.  In baroclinic acoustic streaming, the $\mathit{O}(\epsilon)$ streaming flow is sufficiently strong to induce $\mathit{O}(1)$ rearrangements of the background temperature and density fields over an $\mathit{O}(1/\epsilon)$ time period.  Thus, in contrast to other studies of acoustic streaming in the presence of inhomogeneous temperature fields (e.g. \cite{Cervenka2017}), it is crucial that the expansions for both $\Theta$ and $\rho$ begin at $\mathit{O}(1)$, as first shown in \cite{Chini2014}.

Owing to the large perturbations to the background density field, the natural frequency of an acoustic mode may evolve in time.  Here, we extend the analysis of \cite{Chini2014} by employing a WKBJ approximation to properly capture this slow temporal evolution.  Specifically, a generic dependent field $f(x,y,t)$ is re-expressed as $f(x,y, \phi ,T)$, where $\phi$ and $T$ are treated as independent variables.  The rapidly-varying phase $\phi$ may be written as
\begin{equation}
\phi (t) = \frac{\Phi (T)}{\epsilon},
\end{equation}
where $\mathrm{d} \Phi / \mathrm{d} T$ is of order unity. We define the instantaneous angular frequency $\omega(T)$,
\begin{equation}
\omega(T) = \frac{\mathrm{d} \phi}{\mathrm{d} t} = \frac{\mathrm{d} \Phi}{\mathrm{d} T},
\end{equation}
and expand $\Phi = \Phi_0 +  \epsilon \Phi_1 + O(\epsilon^2)$, so that 
\begin{equation}
\omega = \omega_0 + \epsilon \omega_1 + O(\epsilon^2).
\end{equation}
Finally, in order to distinguish the streaming flow from the acoustic wave, we introduce the fast-time average of a function $f(x,y,\phi,T)$,
\begin{equation}
\bar{f}(x,y,T) = \frac{1}{2n\pi} \int_\phi^{\phi + 2n\pi} f(x,y,s,T) \mathrm{d}s
\end{equation}
for sufficiently large positive integer $n$, so that any function can be decomposed such that
\begin{equation}
f(x,y,\phi,T) = \bar{f}(x,y,T) + f'(x,y,\phi,T),
\end{equation}
where $\overline{f'} = 0$. Thus, $f'$ represents the acoustic wave and $\bar{f}$ the streaming flow.

\subsection{Leading-order multiscale wave/mean-flow interaction equations}

The multiple time-scale governing equations for the coupled acoustic-wave/streaming-flow system were first derived in \cite{Chini2014} and are reproduced here for completeness. The evolution of the streaming fields is governed by the following equations:
\begin{eqnarray}
\bar{\rho}_0 \left( \partial_T \bar{u}_1+\bar{u}_1 \partial_x \bar{u}_1+\bar{v}_1 \partial_y \bar{u}_1  \right) &=& - \frac{\partial_x \bar{\pi}_2}{\gamma}  - \partial_x \left( \bar{\rho}_0 \overline{u_1'^2} \right) - \partial_y \left( \bar{\rho}_0 \overline{u_1' v_1'} \right) + \frac{ \partial_{yy} \bar{u}_1}{Re_s h^2},\quad{}\label{SF:momentum}\\
\partial_y \bar{\pi}_2 &=& 0,\label{SF:momentum_y}\\
\partial_T \bar{\rho}_0 + \partial_x (\bar{\rho}_0 \bar{u}_1) + \partial_y (\bar{\rho}_0 \bar{v}_1) &=& 0, \label{SF:conservation_mass}\\
\partial_T \bar{\Theta}_0 +  \bar{u}_1 \partial_x \bar{\Theta}_0 + \bar{v}_1 \partial_y \left(  \bar{\Theta}_0 + T_B \right) &=& (1-\gamma) (\bar{\Theta}_0 + T_B) (\partial_x \bar{u}_1 + \partial_y \bar{v}_1)+ \frac{\gamma \partial_{yy} \bar{\Theta}_0}{Pe_s h^2 \bar{\rho}_0},\label{SF:heat}\\
\bar{\rho}_0 &=& \frac{1}{\bar{\Theta}_0+T_B}. \label{SF:state}
\end{eqnarray}
%
%
%In this slow-time system, the sole -- but crucial -- effect of the waves arises from the Reynolds stress divergence in the mean $x$-momentum equation (\ref{SF:momentum}) that drives the streaming flow.  The equations governing the leading-order acoustic wave dynamics are
%
In this slow-time system, the sole -- but crucial -- effect of the waves arises from the Reynolds stress divergence in the mean $x$-momentum equation (\ref{SF:momentum}) that drives the streaming flow.  For a homogeneous fluid, this wave-induced force can be offset by a pressure gradient in the bulk:  a Rayleigh streaming flow is driven at \emph{next order} in $\epsilon$ by the Reynolds stress divergence acting in the oscillatory boundary layers that arise near the no-slip channel walls. In contrast, for baroclinic acoustic streaming, the mean temperature gradient causes this wave-induced forcing to be rotational even within the bulk interior of the domain. Consequently, the bulk force that is created cannot be balanced by a mere adjustment of the mean pressure gradient and instead induces a (strong) mean flow even in the absence of diffusive boundary layers. (Further discussion of this distinction is given in \S~\ref{S5}.)

To evaluate this force, we employ the equations governing the leading-order acoustic wave dynamics, \emph{viz.}
\begin{equation}
\omega_0 \bar{\rho_0}  \partial_\phi  u_1'  + \frac{1}{\gamma} \partial_x \pi_1' = 0,\;\;\;\;
%\label{waves:momentum_x}
\partial_y \pi_1' = 0,\label{waves:momentum_xy}\\[-0.05in]
\end{equation}
\begin{equation}
\omega_0 \partial_\phi \rho_1' + \partial_x (\bar{\rho}_0 u_1' ) + \partial_y (\bar{\rho}_0 v_1') = 0,\label{wave:conservation_mass}
\end{equation}
\begin{equation}
\omega_0 \partial_\phi \Theta_1' +  u_1' \partial_x \bar{\Theta}_0 + v_1' \partial_y \left(  \bar{\Theta}_0 + T_B \right) + (\gamma -1) (\bar{\Theta}_0 + T_B) (\partial_x u_1' + \partial_y v_1')=0,\label{wave:temperature}
\end{equation}
\begin{equation}
\pi_1' - \rho_1' (\bar{\Theta}_0 + T_B) - \bar{\rho}_0 \Theta_1' =0.\label{waves:state}
\end{equation}
These equations describe the fast dynamics of approximately linear non-dissipative acoustic waves in a medium whose mean density field $\bar{\rho}_0$ evolves slowly in time.  Conversely, the evolution of $\bar{\rho}_0(x,y,T)$ depends on the acoustic waves, as can be gleaned from inspection of (\ref{SF:momentum})--(\ref{SF:state}).   
%\textbf{As will be discussed in \S~\ref{S5}, systems exhibiting this \emph{quasilinear} (QL) structure increasingly are being used to model anisotropic turbulent shear flows arising in engineering and geophysical applications. In the present investigation, however, the QL dynamical structure arises from a self-consistent asymptotic rather than an \emph{ad hoc} procedure.}  
Note further that, here, unlike in Rayleigh streaming, the oscillatory Stokes layers are \emph{dynamically passive} because the streaming induced by near-wall fluctuating viscous torques is $\mathit{O}(\epsilon^2)$ while the streaming flow governed by (\ref{SF:momentum})--(\ref{SF:state}) is $\mathit{O}(\epsilon)$.  Therefore, the leading-order wave field is required to satisfy only a zero normal-flow boundary condition at each wall, and the details of the oscillatory flow within the Stokes (boundary) layers do not have to be determined at this order (see \S~\ref{S5}). 
%for additional discussion of this point).

Taken together, these two sets of equations form a closed but \emph{two time-scale} system.  As emphasized in \cite{Chini2014}, the fully two-way coupling between the waves and mean flow captured by this multiscale system renders it fundamentally distinct from classical Rayleigh streaming theory, in which the acoustic wave field can be computed first and then the response of the streaming flow to the acoustic wave forcing self-consistently determined (i.e. one-way coupling).  Subsequently, \cite{Karlsen2016, Karlsen2017} also noted this two-way coupling in their computational studies of acoustic streaming in inhomogeneous fluids.  Nevertheless, to make analytical progress, \cite{Chini2014} considered a small Prandtl-number limit in which the coupling is effectively one way.  A primary contribution of the present investigation is to extend the analysis of \cite{Chini2014} to systematically treat fully two-way wave/mean-flow interactions.  This extension requires the derivation of a novel amplitude equation governing the slow evolution of the acoustic waves, which can only be determined by carrying the asymptotic analysis to next order and imposing an appropriate solvability condition.  Heuristically, the slow evolution of the waves is controlled by higher-order terms that, e.g., account for energy exchanges with the streaming flow or with the solid boundaries.  In the next section, we show how these effects can be self-consistently incorporated.
%However, the evolution of the waves on the slow time scale relies on terms that are of the next order, as for instance exchange of energy with the streaming flow or with the solid boundaries. In the next section, we show how they can be taken into account via a solvability condition.
	
%\section{Fast and slow acoustic-wave dynamics \label{S2}}
%\section{Acoustic wave amplitude equation \label{S2}}
\section{Averaging over fast wave dynamics\label{S2}}

We now characterize the dynamics of the acoustic waves on both the fast and slow time scales with the aim of eliminating the need to explicitly simulate the fast evolution.  Inspection of (\ref{waves:momentum_xy})--(\ref{waves:state}) reveals that the wave dynamics directly depends on a single slowly-varying field:  the leading-order mean density $\bar{\rho}_0 = ( \bar{\Theta}_0 + T_B )^{-1}$.  For purposes of the analysis described in this section, this field is presumed to be given.  Consequently, the fluctuation equations (\ref{waves:momentum_xy})--(\ref{waves:state}) comprise a linear homogeneous system, and we henceforth consider a single eigenvector.  A generic fluctuation field $f_1'$ can be expressed as
\begin{equation}
f_1'(x,y,\phi,T)= \frac{A(T)}{2} \left(\hat{f}_1 (x,y,T) e^{i\phi} + \mathrm{c.c.} \right),\label{wave:decomposition}
\end{equation}
where $f_1'$ stands for any fluctuation variable ($u_1'$, $v_1'$, $\rho_1'$, $\pi_1'$,  $\Theta_1'$); $A(T)$ is the slowly-evolving modal amplitude (here taken to be real without loss of generality); $\hat{f}$ is a complex function that describes the spatial structure of the mode; and c.c. denotes the complex conjugate.  A normalization condition, specified subsequently, must be imposed on $\hat{f}$ to render this decomposition unique.  We next describe the determination of the spatial structure of the mode (as defined by the functions $\hat{u}_1$, $\hat{v}_1$, etc.) and then derive a novel amplitude equation governing the slow evolution of the generally \emph{a priori} unknown function $A(T)$.

\subsection{Mode structure \label{S2_1}}

Substituting the decomposition (\ref{wave:decomposition}) into the fluctuation equations (\ref{waves:momentum_xy})--(\ref{waves:state}) yields a linear but non-separable two-dimensional (partial) differential eigenvalue problem for the spatial structure and frequency $\omega_0$ of the leading-order fluctuation fields.  By continuing to exploit the small aspect-ratio limit, we nevertheless are able to reduce the required computation to the solution of a one-dimensional eigenvalue problem -- a crucial simplification.

To proceed, we note that, using  (\ref{wave:decomposition}), (\ref{wave:conservation_mass})--(\ref{waves:state}) can be combined to deduce
\begin{equation}
\hat{\pi}_1 = \frac{i \gamma}{\omega_0} \left(\partial_x \hat{u}_1 + \partial_y \hat{v}_1 \right).
\end{equation}
With this expression for the acoustic wave pressure $\hat{\pi}_1$, the momentum equations (\ref{waves:momentum_xy}) become two coupled partial differential equations for $\hat{u}_1$ and $\hat{v}_1$:
\begin{equation}
\partial_x \left( \partial_x \hat{u}_1 + \partial_y \hat{v}_1 \right) + \omega_0^2 \bar{\rho}_0 \hat{u}_1 =0 ,\label{wave:eq1}
\end{equation}
\begin{equation}
\partial_y \left( \partial_x \hat{u}_1 + \partial_y \hat{v}_1 \right)=0.\label{wave:eq2}
\end{equation}
A further reduction to a single ordinary differential equation is possible by formally integrating a linear combination of these equations,
\begin{equation}
\partial_y (\ref{wave:eq1}) + \partial_x (\ref{wave:eq2}) \Rightarrow \partial_y \left( \bar{\rho}_0 \hat{u}_1  \right) = 0 \Rightarrow \hat{u}_1 = \frac{q}{\bar{\rho}_0}, \label{wave:H}
\end{equation}
where $q$ is an unknown function of $x$ and $T$ only. Integration of (\ref{wave:eq2}) gives
\begin{equation}
\partial_x \hat{u}_1 + \partial_y \hat{v}_1 = g, \label{wave:G}
\end{equation}
where $g$ is a second unknown function of $x$ and $T$ only. Equations (\ref{wave:eq1}), (\ref{wave:H}) and (\ref{wave:G}) imply that $q = - g' / \omega_0^2$, where a prime is used to denote partial differentiation of the function $g$ with respect to $x$, since the $T$ dependence is parametric.  The general solution of this system of equations can be obtained using the kinematic boundary condition $\hat{v}_1(x,y=0,T) = 0$, \emph{viz.}
\begin{equation}
\hat{u}_1 = - \frac{g'}{\omega_0^2 \bar{\rho}_0},\label{wave:uhat}
\end{equation}
\begin{equation}
\hat{v}_1 = yg + \partial_x \left(\frac{g'}{\omega_0^2}  \int_0^y \frac{\mathrm{d}y}{\bar{\rho}_0}  \right) .\label{wave:vhat}
\end{equation}
Finally, the upper boundary condition $\hat{v}_1(x,y=1,T) = 0$ provides a constraint on $g$ and $\omega_0$ in the form of the ordinary differential eigenvalue problem
\begin{equation}
\frac{\mathrm{d}}{\mathrm{d}x} \left(  g' \alpha \right)  + \omega_0^2 g = 0,\label{wave:ode_g}
\end{equation}
where
\begin{equation}
\alpha(x,T) = \int_0^1 \frac{\mathrm{d}y}{\bar{\rho}_0}. \label{alpha}
\end{equation}

%%The ability to describe a two-dimensional acoustic field with a single ordinary differential equation is a result of the thin layer approximation: in any container of aspect ratio of order unity, the eigenmodes would have to be found through a two-dimensional eigenvalue problem. This provides an important simplification to the analysis of these acoustic modes, and we thereafter characterize this function $g$.

To characterize the function $g(x,T)$, we first take, without loss of generality, $\hat{u}_1$ to be a real field: (\ref{wave:eq1}) and (\ref{wave:eq2}) then imply that $\hat{v}_1$ and, thence, $g$ are also real-valued fields. 
%Moreover, in order to ensure zero mass exchange across $x=0$, 
Moreover, in order to ensure $\hat{u}_1(x=0,y,T)=0$, the ordinary differential equation (\ref{wave:ode_g}) must be solved subject to the boundary conditions $g'(0) = g'(2\pi ) = 0$. This requirement leads to an orthogonality condition: let $(g_A,g_B)$ be two eigenvectors and $(\omega_A, \omega_B)$ their angular eigenfrequencies; then
\begin{equation}
\int_0^{2\pi} g_A(x) g_B(x) \mathrm{d}x = \frac{1}{\omega_A^2 - \omega_B^2} \left[  \alpha (g_Ag'_B - g_B g'_A) \right]_0^{2\pi} = 0.\label{g:orthogonality}
\end{equation}
Equation (\ref{g:orthogonality}) provides a convenient scalar product on eigenvectors, and therefore we normalize them according to 
\begin{equation}
\int_0^{2\pi}g(x)^2 \mathrm{d}x = 1. \label{wave:normalization}
\end{equation}
This normalization condition resolves the ambiguity in the definition of $A(T)$ and $\hat{f}_1$ in (\ref{wave:decomposition}).
\subsection{Wave amplitude\label{S2_2}} 
As explained in the previous subsection, the acoustic mode shape $g(x,T)$ and frequency $\omega_0(T)$ can be computed at every time $T$ for a given mean density profile $\bar{\rho}_0(x,y,T)$. The advection of hot or cold gas by the streaming flow will cause both $g$ and $\omega_0$ to evolve on the slow time scale and induce a two-way coupling between the waves and the streaming flow. The amplitude $A(T)$ of the acoustic mode is also expected to evolve on this slow time scale owing to dissipation by viscosity, energy exchanges with the streaming flow and walls and/or external forcing. To obtain an evolution equation for $A(T)$, we proceed as follows, relegating details to the appendices.\\
%

%\begin{enumerate}
 
\noindent
1. We collect terms in the dimensionless governing equations at $\mathit{O}(\epsilon^2)$. The resulting equations are reported in Appendix~\ref{appA}.\\

\noindent
2.  We make the ansatz that generic $\mathit{O}(\epsilon^2)$ fluctuation field $f_2'$ can be represented as
\begin{equation}
f_2'(x,y,T,\phi) = \frac{B(T)}{2} \left(\hat{f}_2 (x,y,T) e^{i\phi} + \mathrm{c.c.} \right)
\end{equation}
and, in direct analogy with the manipulations performed in \S~\ref{S2_1}, reduce the $\mathit{O}(\epsilon^2)$ fluctuation system for the five unknown fields $\hat{u}_2$, $\hat{v}_2$, $\hat{\pi}_2$, $\hat{\rho}_2$ and $\hat{\Theta}_2$ to a system of two equations for the two fields $\hat{u}_2$ and $\hat{v}_2$. [Harmonics of the form $e^{i2\phi}$ also exist at this order, but are non-resonant and thus do not contribute to the slow-time dynamics; accordingly, these harmonics need not be explicitly computed.]  The resulting system of equations has the form
\begin{equation}
\partial_x \left( \partial_x \hat{u}_2 + \partial_y \hat{v}_2 \right) + \omega_0^2 \bar{\rho}_0 \hat{u}_2 =\mathcal{F} ,\label{wave_2:eq1}
\end{equation}
\begin{equation}
\partial_y \left( \partial_x \hat{u}_2 + \partial_y \hat{v}_2 \right)=\mathcal{G}.\label{wave_2:eq2}
\end{equation}
The linear operator acting on the left-hand side of this system is identical to that arising in the leading-order fluctuation equations (\ref{wave:eq1})--(\ref{wave:eq2}).  The right-hand side functions $\mathcal{F}$ and $\mathcal{G}$ include resonant forcing terms involving the leading-order fluctuations fields ($f_1'$). Analytical expressions for the imaginary parts of $\mathcal{F}$ and $\mathcal{G}$, which are needed for the derivation of the (real) amplitude equation, are given in Appendix~\ref{appB}.\\

\noindent
3. We derive a solvability condition for the $\mathit{O}(\epsilon^2)$ system, which requires determination of the adjoint linear operator, by invoking the Fredholm alternative theorem in the usual manner; see Appendix~\ref{appC}.\\

\noindent
4. We enforce the solvability condition to obtain an equation for $\mathrm{d}A/\mathrm{d}T$; cf. Appendix~\ref{appD}.\\
%\end{enumerate}

\noindent
Employing this procedure, we derive the following novel amplitude equation:
\begin{eqnarray}
\frac{2}{A\omega_0^{-1}}&& \frac{\mathrm{d}(A\omega_0^{-1})}{\mathrm{d}T} = -\frac{i\omega_0}{Pe_s h^2} \int_D \mathrm{d}x \mathrm{d}y\,g \partial_{yy} \hat{\Theta}_1- \frac{1}{2\omega_0^2} \int_D \mathrm{d}x\mathrm{d}y\,g'^2\left( \bar{u}_1\partial_x \bar{\rho}_0^{-1}+ \bar{v}_1 \partial_y \bar{\rho}_0^{-1} \right)\nonumber\\
&& +\,\int_D \mathrm{d}x\mathrm{d}y\, (\partial_x \bar{u}_1 + \partial_y \bar{v}_1) \left[(1-\gamma ) g^2  + \frac{g'^2}{\omega_0^2\bar{\rho}_0}\left(  \frac{Pe_s}{Re_s} - \frac{1}{2}\right)   \right],\label{wave:amplitude_equation}
\end{eqnarray}
where $\int_D$ refers to definite integration over the spatial domain.  (Note that, since the temperature and velocity fluctuations are out of phase, $\hat{\Theta}_1$ is strictly imaginary.)  The lack of terms nonlinear in $A$ in (\ref{wave:amplitude_equation}) confirms that phenomena such as shock-wave formation or harmonic generation are sub-dominant dynamical processes in the given parameter regime relative to, for example, heat exchange with the boundaries or energy exchange with the evolving stratified environment.  In particular, the first term on the right-hand side of (\ref{wave:amplitude_equation}) accounts for the time-mean heat transfer between the waves and the boundaries.  In \S~\ref{sec_stability}, we demonstrate that this term can be positive:  in this scenario, the waves are driven by a process that is loosely akin to the classical thermoacoustic instability in which acoustic waves in a channel can be excited when a temperature gradient is imposed \emph{along} the channel walls \citep{Swift1988}.  Owing to the occurrence of the mean fields $\bar{u}_1$, $\bar{v}_1$ and $\bar{\rho}_0$ in (\ref{wave:amplitude_equation}), the amplitude equation is, in fact, nonlinear.  A distinguishing feature of (\ref{wave:amplitude_equation}) is that, unlike amplitude equations derived in numerous other contexts, determination of the coefficients requires evaluation of \emph{functional derivatives} that capture the $\mathit{O}(1)$ variations in generic eigenfunction $\hat{f}_1$ caused by changes in the mean density field $\bar{\rho}_0$ that occur on the slow time $T$; see Appendices~\ref{appB} and~\ref{appD} for details.

The quantity $A\omega_0^{-1}$ on the left-hand side of (\ref{wave:amplitude_equation}) is proportional to the square-root of the dimensionless energy of the acoustic wave. Indeed, the leading-order dimensionless kinetic energy $E_K$ of the acoustic wave averaged over the fast time scale is given by
\begin{equation}
\overline{E}_K \equiv \frac{1}{2} \int_0^{2\pi} \mathrm{d}x \int_0^1 \mathrm{d}y \bar{\rho}_0 \overline{u_1'^2} = \frac{A(T)^2}{4} \int_0^{2\pi} \mathrm{d}x \int_0^1 \mathrm{d}y \bar{\rho}_0 \hat{u}_1^2.
\end{equation}
This expression can be evaluated using (\ref{wave:uhat}) to obtain
\begin{equation}
\overline{E}_K = \frac{A(T)^2}{4\omega_0^4} \int_0^{2\pi} \mathrm{d}x \int_0^1 \mathrm{d}y \frac{g'^2}{\bar{\rho}_0} =  \frac{A(T)^2}{4\omega_0^4} \int_0^{2\pi} \mathrm{d}x  g'^2 \alpha.
\end{equation}
The last integral reduces to $\omega_0^2$ following an integration by parts and ulitisation of the differential equation (\ref{wave:ode_g}) and the normalization condition (\ref{wave:normalization}), yielding
\begin{equation}
\overline{E}_K = \left( \frac{A(T)}{2 \omega_0} \right)^2.
\end{equation}
The amplitude equation therefore can be interpreted as an energy balance for the acoustic wave, with the left-hand side of (\ref{wave:amplitude_equation}) equalling $(1/\overline{E}_K) \mathrm{d}\overline{E}_K / \mathrm{d}T$.

An important outcome of this study is that even allowing for two-way wave/mean-flow coupling, the WKBJ analysis enables the streaming flow to be computed \emph{without} evolving the sound waves over the fast time scale.  (Of course, in the absence of two-way coupling, as in classical Rayleigh streaming, this averaging is trivial.)  Instead, the evolving spatial structure of the waves can be computed at every coarse time step (in a numerical simulation) by solving the one-dimensional eigenvalue problem (\ref{wave:ode_g}), while the evolution of the amplitude can be determined by integrating (\ref{wave:amplitude_equation}) over the slow time scale.  These computations are performed in conjunction with the numerical solution of the streaming equations (\ref{SF:momentum})--(\ref{SF:state}).

%%\section{Limit of one-way coupling \label{S3}}
%\section{Analytical streaming solution in the weak-wave limit\label{S3}}
\section{One-way coupling\label{S3}}

Although the asymptotically-reduced equations constitute a substantial simplification of the full compressible Navier--Stokes equations, they defy analytical solution owing to the occurrence of nonlinearities and two-way coupling between the waves and the streaming flow.  For sufficiently weak streaming, however, an approximate steady-state solution can be derived: in this limit, mean advection is weak and the mean density perturbations are small, thereby ameliorating these two difficulties.  Here, we derive this approximate analytical solution and demonstrate that it accurately describes results obtained from direct numerical simulations of the compressible Navier--Stokes and heat equations reported in the literature.

\subsection{Acoustic waves\label{S3_ac}}

In this section, we assume that the mean density profile varies little and thus can be accurately approximated by the diffusive solution (\ref{Background_fields}), i.e.
\begin{equation}
\bar{\rho}_0 = \frac{1}{T_B} =\frac{1}{1 + \Gamma y}.
\end{equation}
The coefficient $\alpha$ defined in (\ref{alpha}) then does not depend on $x$ and is simply equal to $1 + \Gamma /2$. The acoustic-wave eigenfunction $g$ characterizing the shape of the acoustic mode follows from the solution of the second-order differential equation (\ref{wave:ode_g}) and, with the prescribed boundary conditions $g'(0)=g'(2\pi)=0$ and the normalization condition (\ref{wave:normalization}), is given by
\begin{equation}
g(x) = \frac{\cos(nx)}{\sqrt{\pi}}, \label{waves:mode1g}
\end{equation}

where the integer $n$ is set to unity for this study. The angular frequency of this mode $\omega_0 = \sqrt{1 + \Gamma /2}$, and the velocity field ($\hat{u}_1$, $\hat{v}_1$) follows from (\ref{wave:uhat})--(\ref{wave:vhat}) and reduces to 
\begin{equation}
\hat{u}_1 = \frac{(1+ \Gamma y) \sin (x)}{(1 + \Gamma /2) \sqrt{\pi}},\label{waves:mode1u}
\end{equation}
\begin{equation}
\hat{v}_1 = \frac{\Gamma y\left(1-y  \right) \cos(x)}{2(1 + \Gamma /2)\sqrt{\pi}} . \label{waves:mode1v}
\end{equation}
\begin{figure}
%  \centerline{\includegraphics{mode1_ac_field}}
%   \centerline{\includegraphics[width=0.9\linewidth]{OneWayCouplingLinFaroukWaveFieldS}}
        \centerline{\includegraphics*[width=0.9\linewidth]{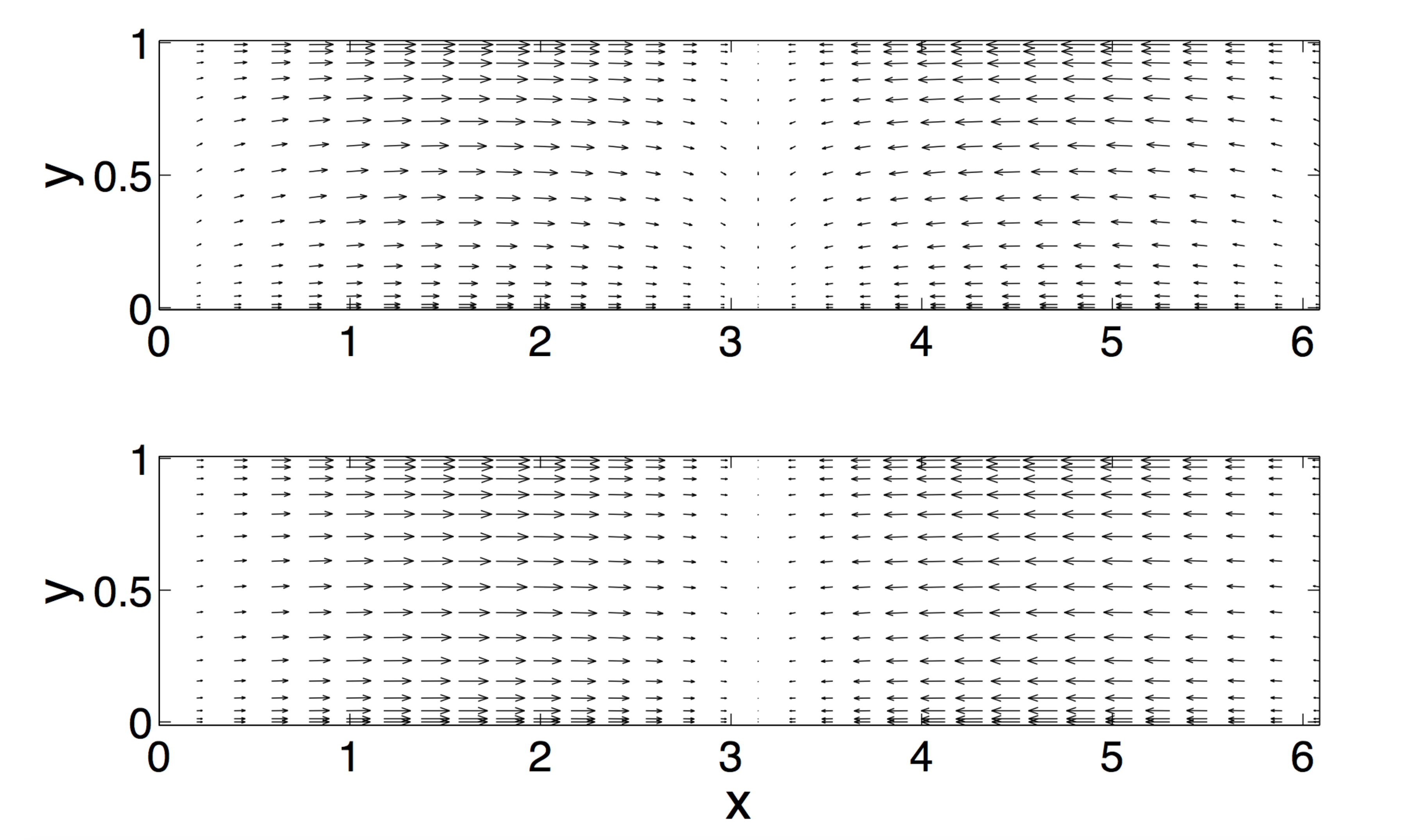}}
  \caption{Standing acoustic-wave velocity field corresponding to the first (i.e. $n=1$) eigenmode for $\Gamma=1$ (upper) and $\Gamma = 0.2$ (lower), with $\gamma=1.4$. Particularly in the upper plot, it is evident that the wall-normal ($y$) velocity component is non-zero owing to the background density stratification and, crucially, the wave is rotational.  The lower plot is included to illustrate the dependence of the wave field on the stratification and because $\Gamma=0.2$ is the value used in the direct numerical simulation performed by \cite{Lin2008} to which we compare our theoretical predictions.}
\label{fig:mode1_ac_field}
\end{figure}

This velocity field is plotted in figure \ref{fig:mode1_ac_field} for two different values of $\Gamma$.  A rotational (vortical) component may be discerned, particularly for the $\Gamma=1$ scenario, even though viscous torques are absent since the dynamics in oscillatory boundary layers has been self-consistently omitted.   Crucially, the Reynolds stress divergence terms, here denoted $R(x,y)$, arising in (\ref{SF:momentum}) and responsible for driving the streaming flow can be explicitly evaluated:
\begin{equation} \label{one_way:RS}
R(x,y) = - \partial_x \left( \bar{\rho}_0 \overline{u_1'^2} \right) - \partial_y \left( \bar{\rho}_0 \overline{u_1' v_1'} \right) = - \frac{A^2}{2\pi (1 + \Gamma /2)^2 } \left( 1 + \frac{\Gamma}{4} + \frac{\Gamma y}{2} \right) \sin (2x).
\end{equation}
If $\Gamma=0$, i.e. for a homogeneous fluid, $R(x,y)= \partial_x [(A^2/4\pi) \cos (2x)]$.  Clearly, in that case, the wave-induced Reynolds-stress divergence can be balanced by a mean pressure gradient, so that streaming is \emph{not} directly driven; instead, the associated Rayleigh streaming flow arises at next order in $\epsilon$ owing to the action of viscous torques within oscillatory boundary layers.  If, however, $\Gamma \neq 0$, then $R(x,y)$ can no longer be reduced to gradient form and, consequently, directly drives a streaming flow.

\subsection{Streaming flow\label{S:streaming_flow}}

We assume that the steady streaming driven by $R(x,y)$ given in (\ref{one_way:RS}) is sufficiently weak that the streaming equations (\ref{SF:momentum})--(\ref{SF:state}) can be linearised.  Note that, unlike Rayleigh streaming, the mean flow is compressible even though the streaming Mach number is negligible.

In a steady state, the pressure field can be eliminated from the linearised versions of equations (\ref{SF:momentum})--(\ref{SF:momentum_y}), yielding
\begin{equation}
\partial_{xy} R = - \frac{\partial_{yyyx} \bar{u}_1}{Re_s h^2},\label{one_way:eq1}
\end{equation}
while conservation of mass and internal energy (\ref{SF:conservation_mass})--(\ref{SF:heat}) imply
\begin{equation}
\partial_x \bar{u}_1 = - \frac{(1 + \Gamma y)}{\Gamma Pe_s h^2} \partial_{yyy} \bar{\Theta}_0.\label{one_way:eq2}
\end{equation}
Finally, we obtain from equations (\ref{one_way:RS})--(\ref{one_way:eq2}) a single differential equation for $\bar{\Theta}_0$: 
\begin{equation}
(1 + \Gamma y ) \partial_{yyyyyy} \bar{\Theta}_0 + 3 \Gamma \partial_{yyyyy} \bar{\Theta}_0 = - \frac{2 A^2 \Gamma^2 Re_s Pe_s h^4}{\pi (2+\Gamma)^2} \cos(2x).\label{one_way:Theta0eqn}
\end{equation}
This equation is supplemented with the boundary conditions:
\begin{enumerate}
\item $\bar{\Theta}_0 (x,y=0) = \bar{\Theta}_0(x,y=1)=0$, since the wall temperatures are held constant; 
\item $\partial_{yy}\bar{\Theta}_0 (x,y=0) = \partial_{yy}\bar{\Theta}_0 (x,y=1) = 0$, to enforce $\bar{v}_1=0$ at the walls [see (\ref{SF:conservation_mass})--(\ref{SF:state})]; and
\item $\partial_{yyy}\bar{\Theta}_0 (x,y=0) = \partial_{yyy}\bar{\Theta}_0 (x,y=1) = 0$, to enforce the no-slip boundary condition at each wall, upon using (\ref{one_way:eq2}).
\end{enumerate}

With these boundary conditions, a unique solution can be found.  For illustration, we focus on the case $\Gamma = 1$, for which
\begin{equation}
\bar{\Theta}_0(x,y) = - \frac{2 A^2 Re_s Pe_s h^4}{9\pi} G(y) \cos(2x), \label{one_way:T1}
\end{equation}
where
\begin{eqnarray}
G(y) =&& \frac{1}{1080 (-3+ \log (16))} \bigg[ 60 (1+y)^2 \log(1+y) \\
&&+\,y (94-222\log(2)-90y-20y^2 -5y^3 (-5+\log(64)) +3 y^4 (-3+\log(16)))  \bigg]. \nonumber
\end{eqnarray}
%This function $G(y)$ is always negative, expressing that acoustic streaming cools down the gas.
For the self-consistency of this approximation, the assumptions of one-way coupling and linear streaming dynamics require that the mean density change little during the evolution, i.e. $\bar{\Theta}_0 \ll 1$, and that the nonlinear terms be negligible, e.g. $\bar{u}_1^2 \ll 1$. From equations (\ref{SF:momentum})--(\ref{SF:state}) and (\ref{one_way:T1}), these constraints imply (for $\Gamma =1$) upper bounds on the dimensionless parameter combinations $A^2 Re_s h^2$ and $A^2 Re_s Pe_s h^4$.

\subsection{Comparison with previous work\label{sec:comp}}

Using the analytical solution (\ref{one_way:T1}), various properties of the streaming flow can be deduced. First, the maximum dimensional baroclinic streaming velocity obtained here
\begin{equation}
%\tilde{u}_B = \left(\mathrm{max}~ |G| \right)  A^2 Re_s h^2 U_*,
\tilde{u}_B \propto \left(\mathrm{max}~ |G| \right)  A^2 Re_s h^2 U_*
\end{equation}
can be compared to the corresponding velocity $\tilde{u}_R$ in Rayleigh streaming resulting from dissipation in the Stokes boundary layers, i.e.
\begin{equation}
%\tilde{u}_R = \frac{3 \epsilon}{8} U_*.
\tilde{u}_R \propto \epsilon A^2 U_*.
\end{equation}
Clearly, for small values of $\epsilon$, the baroclinic streaming dominates any streaming driven by viscous torques acting in near-wall Stokes layers.

\begin{figure}
%  \centerline{\includegraphics{mode1_mean_velocity}}
%    \centerline{\includegraphics[width=0.9\linewidth]{OneWayCouplingLinFaroukMeanFields_SHADING.pdf}}
%     \centerline{\includegraphics[width=0.9\linewidth]{OneWayCouplingLinFaroukMeanFields_SHADINGv3.pdf}}
          \centerline{\includegraphics[width=0.9\linewidth]{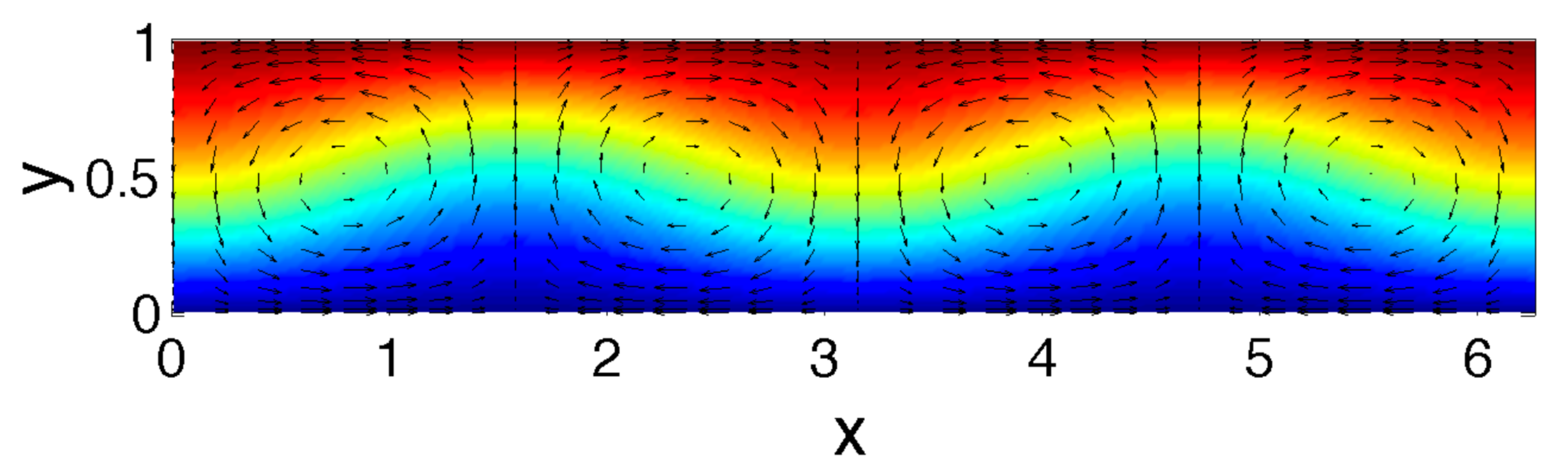}}
  \caption{Analytically-predicted baroclinic streaming flow obtained in the limit of one-way coupling (i.e. for sufficiently weak sound waves, which implies upper bounds on $A^2Re_sh^2$ and $A^2Re_sPe_sh^4$).  Colour is used to indicate the total temperature  (blue is cold).  Recall that the lower boundary is colder than the upper boundary, and that gravitational forces have been neglected.  The parameter values are the same as those used to compute the sound wave shown in figure~\ref{fig:mode1_ac_field} (lower plot); additionally, $A=6$, $h=2.3$, $Re_s=5.7$ and $Pe_s=4.1$.  Collectively, these parameters correspond closely to the values employed by \cite{Lin2008} in their full numerical simulations of stratified acoustic streaming.  In contrast to Rayleigh streaming, the cells span the channel.}
  \label{fig:mode1_mean_velocity}
\end{figure}

The streaming velocity field ($\bar{u}_1$,  $\bar{v}_1$) computed from (\ref{one_way:Theta0eqn}) for $\Gamma=0.2$ is plotted along with the total temperature $T_B+\bar{\Theta}_0$ in figure~\ref{fig:mode1_mean_velocity}.  Clearly, baroclinic streaming differs from Rayleigh streaming not only in intensity but also in spatial structure; specifically, here the streaming cells span the channel while in Rayleigh streaming the cells are stacked in the wall-normal direction.  These various distinguishing properties accord with prior experiments and numerical simulations; e.g. see \cite{Loh2002, Hyun2005, Lin2008, Nabavi2008, Aktas2010, Dreeben2011}.

Quantitative comparisons can be made with the results of \cite{Lin2008}, who performed direct numerical simulations of the compressible Navier--Stokes and heat equations specifically to investigate the impact of acoustic streaming in a thin channel on cross-channel heat transport.  Thus, the system they considered is very similar to that studied here.  In the absence of thermal driving, the streaming flow is accurately predicted by using Rayleigh's formulation and, accordingly, exhibits a pattern of counter-rotating cells \emph{stacked} in the $y$ direction.  When a temperature difference is imposed, however, the stacked cells merge, resulting in counter-rotating cells that span the channel.  In particular, for their case 1C, corresponding to the largest imposed temperature differential, the dimensionless parameters used by \cite{Lin2008} are approximately
\begin{equation}
\epsilon = 10^{-2},~~~ \gamma = 1.4, ~~~ \Gamma = 0.2, ~~~ h = 2.3, ~~~ Re_s = 5.7, ~~~ Pe_s = 4.1.
\end{equation}
The amplitude $A$ of the acoustic waves is not reported, but can reasonably be assumed to be similar to the value arising in the absence of thermal driving, for which $A \approx 6$.
The authors report the values of the $x$ (resp. $y$) component of the dimensional streaming velocity at $x=3\pi/4$ (resp. $x= \pi /2$). We compute $\bar{u}_1$ and $\bar{v}_1$ for the same parameters in this one-way coupling limit, and then obtain the corresponding dimensional velocities by multiplying $\bar{u}_1$ by the sound speed $a_* = 353~\mathrm{m} \cdot \mathrm{s}^{-1}$ and $\bar{v}_1$ by $\sqrt{\epsilon}h a_*$.
\begin{figure}
  \centerline{\includegraphics[scale=0.75]{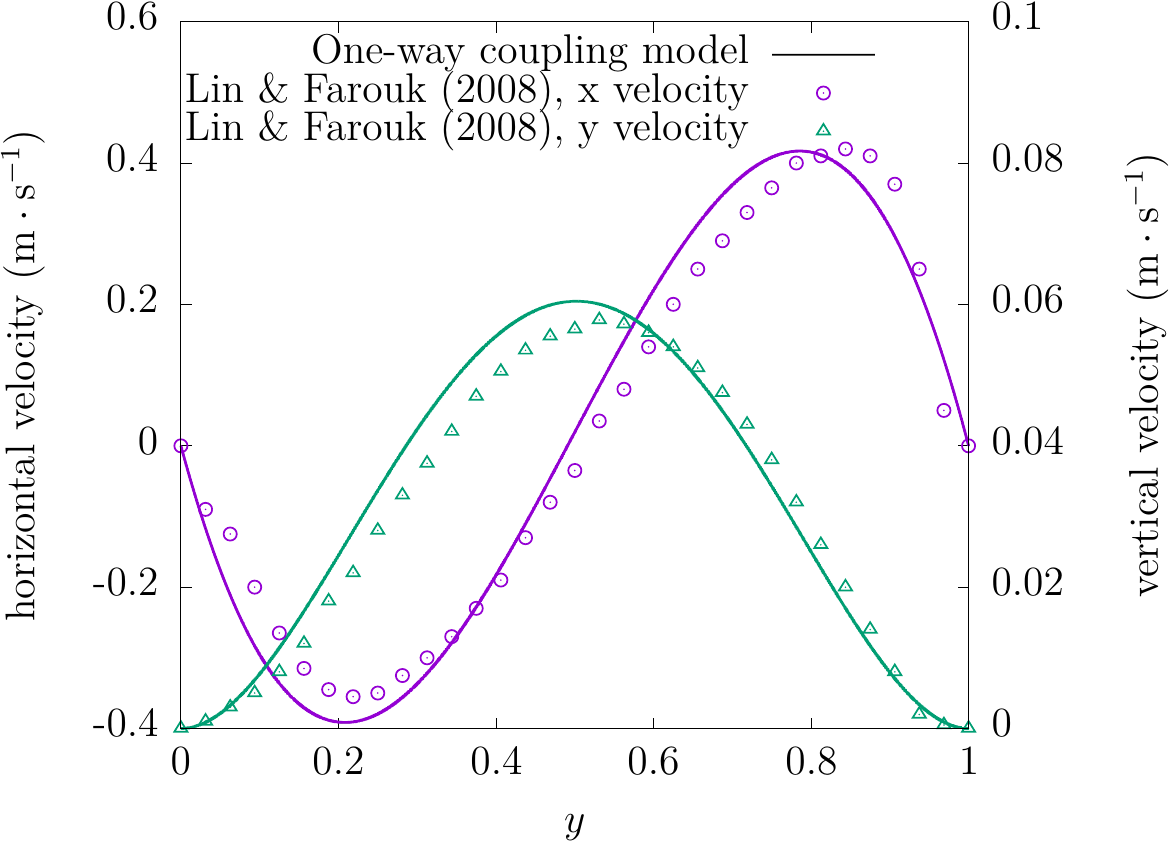}}
%    \centerline{\includegraphics[scale=0.75]{comp_lin2008-2.pdf}}
  \caption{Comparison of the $x$ and $y$ components of the streaming velocity field extracted from the numerical simulations of the instantaneous (compressible) Navier--Stokes and heat equations performed by \cite{Lin2008} (their case 1C) with the corresponding streaming velocities predicted from the present theory in the limit of one-way wave/mean-flow coupling and linear streaming dynamics.  The parameters correspond to:  $A=6$, $\Gamma = 0.2$, $\gamma = 1.4$, $h = 2.3$, $Re_s = 5.7$ and $Pe_s = 4.1$.  Excellent quantitative agreement is observed \emph{without} the use of any fitting parameters.}
\label{fig:comp_lin2008}
\end{figure}
Figure~\ref{fig:comp_lin2008} shows the resulting comparison.  The evident quantitative agreement -- with \emph{no} adjustable parameters and despite the fact that the numerical simulations of \cite{Lin2008} include several physical effects (oscillatory boundary-layer dynamics, viscous heating, inertia and temperature-dependent viscosity and diffusivity coefficients) not incorporated in our analysis -- provides strong confirmation of the baroclinic streaming theory developed by \cite{Chini2014} and systematically extended in the present study.
\subsection{Heat flux enhancement}

The streaming flow enhances cross-channel heat transport, as we now demonstrate using the leading-order solutions derived in this section.  For this purpose, we introduce the steady-state Nusselt number $Nu$ as the ratio of the (dimensional) total heat flux $\dot{Q}_*$ to the diffusive flux,
\begin{equation}
Nu = \frac{\dot{Q}_*}{2\pi \kappa \Delta \Theta_* (k_* H_*)^{-1}},
\end{equation}
and compute $Nu-1$. The total heat flux $\dot{Q}_*$ is evaluated at the top boundary,
\begin{equation}
\dot{Q}_* =  \int_0^{2\pi k_*^{-1}} \kappa \partial_{\tilde{y}} \tilde{T}(\tilde{x}, \tilde{y}=H_*) \mathrm{d}\tilde{x} = \kappa (k_*H_*)^{-1} T_* \left[ \int_0^{2\pi} \partial_y \left(T_B +  \bar{\Theta}_0 \right) (x, y=1)  \mathrm{d}x +O(\epsilon) \right],
\end{equation} 

so that the Nusselt number is given at leading order by
\begin{equation}
Nu = 1+ \frac{1}{2\pi \Gamma} \int_0^{2\pi} \partial_y \bar{\Theta}_0(x,y=1) \mathrm{d}x + O(\epsilon).
\label{eq:heat_flux_def}
\end{equation}
Since viscous heating is neglected, the top and bottom heat fluxes are equal in a steady state; hence, the following result holds:
\begin{eqnarray}
\int_0^{2\pi} \left[T \partial_y \Theta \right]_0^1 \mathrm{d}x =&& \int_0^{2\pi} \big[ (1+ \Gamma) \partial_y \Theta (x,y=1) - \partial_y \Theta (x,y=0) \big] \mathrm{d}x,\\
=&& \Gamma \int_0^{2\pi} \partial_y \bar{\Theta}_0 (x,y=1) \mathrm{d}x + O(\epsilon).
\end{eqnarray}
Moreover, from equations (\ref{SF:conservation_mass})--(\ref{SF:state}), the following equality can be derived,
\begin{equation}
\int_0^{2\pi} \mathrm{d}x \int_0^1 \mathrm{d}y\,T \partial_{yy} \bar{\Theta}_0=0 \Rightarrow \int_0^{2\pi} \left[T \partial_y \Theta \right]_0^1 \mathrm{d}x = \int_0^{2\pi} \mathrm{d}x \int_0^1 \mathrm{d}y \,(\partial_y T)(\partial_y \bar{\Theta}_0 )+ O(\epsilon).
\end{equation}
Finally, using the boundary condition $\bar{\Theta}_0=0$ at the top and bottom walls, we obtain 
\begin{equation}
Nu -1 = \frac{1}{2\pi \Gamma^2} \int_0^{2\pi} \mathrm{d}x \int_0^1 \mathrm{d}y (\partial_y T) (\partial_y \bar{\Theta}_0) + O(\epsilon) =  \frac{1}{2\pi \Gamma^2} \int_0^{2\pi} \mathrm{d}x \int_0^1 \mathrm{d}y (\partial_y \bar{\Theta}_0)^2 + O(\epsilon).
\end{equation}
In the limit of one-way coupling and for $\Gamma = 1$, we evaluate this quantity with the approximate expression (\ref{one_way:T1}) for $\bar{\Theta}_0$ derived in \S~\ref{S:streaming_flow},
\begin{equation} \label{one_way:heat_flux}
Nu -1 \simeq 3.2\times 10^{-10}(A^2 Re_s Pe_s h^4)^2.
\end{equation}
This expression should be contrasted with that obtained for Rayleigh streaming (i.e. for sufficiently small temperature differences) and computed by \cite{Vainshtein1995}:
\begin{equation}
%\left( Nu -1 \right)_{\mathrm{R}} = 6.2\times 10^{-6} \left(\epsilon Pe_s h^2 \right)^2.
\left( Nu -1\right)_{\mathrm{R}} = 6.2  \times 10^{-5} \left(\epsilon A^2 Pe_s h^2\right) ^2.
\end{equation}

This comparison indicates that, although the Nusselt number derived here is very small (as noted in \S~\ref{S:streaming_flow}, for the validity of the one-way coupling assumption the dimensionless parameter combination $A^2 Re_s Pe_s h^4$ cannot be too large), it nevertheless is orders of magnitude larger than $Nu$  resulting from boundary-layer-driven acoustic streaming provided that, numerically, $\epsilon\ll 10^{-2}$.  Equation~(\ref{one_way:heat_flux}) also suggests that significant heat transport enhancement may be achieved in the limit $A^2 Re_s Pe_s h^4 \gg 1$, i.e. when there is strong two-way coupling between the acoustic waves and the streaming flow. This strong coupling scenario is investigated in \S~\ref{S4}.

\subsection{Stability of the quiescent background state\label{sec_stability}}
Intriguingly, the amplitude equation (\ref{wave:amplitude_equation}) suggests that an acoustic wave may be amplified via interaction with the solid boundaries. Indeed, the first term on the right-hand side of (\ref{wave:amplitude_equation}) does not depend on the streaming flow, and quantifies how the divergence of the acoustic-wave velocity field (i.e. $g(x)$) and the net heat flux to/from the solid boundaries [proportional to $\partial_y \hat{\Theta}_1 (x,y=1) - \partial_y \hat{\Theta}_1 (x,y=0)$] are coupled.  In a certain parameter regime, this system may act as a thermoacoustic engine and spontaneously generate an acoustic wave that, in turn, would drive a streaming flow and increase the cross-channel heat transport. 

To see this, note that $\hat{\Theta}_1$ can be determined from the acoustic wave solution derived in \S~\ref{S3_ac} and then used to evaluate
\begin{equation}
-\frac{i\omega_0}{Pe_s h^2} \int_D \mathrm{d}x \mathrm{d}y\,g  \partial_{yy} \hat{\Theta}_1 = \frac{-2 \Gamma \left[1+(2-\Gamma) (\gamma -1) \right]}{Pe_s h^2 (1 + \Gamma /2)}.
\end{equation}
Since $\gamma > 1$, this term is negative and damps the acoustic waves, \emph{unless} the thermal forcing $\Gamma$ is sufficiently strong. In that case, a proper discussion of the stability of the quiescent background state would require the evaluation of the acoustic-wave energy dissipation and temperature profile in the oscillatory boundary layers, which has been neglected here (see \S~\ref{S5}). Nevertheless, this calculation suggests that such an instability, in which an acoustic wave spontaneously grows by a thermoacoustic effect even in the absence of an imposed wall-parallel ($x$) temperature gradient, is possible.

%\section{Numerical simulation of the fully coupled wave/mean-flow system\label{S4}}
\section{Two-way coupling\label{S4}}

The analytical solution derived in \S~\ref{S3} and used to compute the heat flux enhancement is not valid when there is two-way coupling between the acoustic waves and the streaming flow, as occurs for sufficiently large values of the dimensionless parameter groups $A^2 Re_s h^2$ and $A^2 Re_s Pe_s h^4$.  To investigate the dynamics in this regime, we perform numerical simulations of the reduced equations derived in \S~\ref{S1} and \S~\ref{S2} using the spectral computing environment Dedalus \citep{Burns}.

The spatial domain is discretized on a $512 \times 128$ ($x\times y$) grid using a Fourier--Chebyshev pseudospectral scheme.  Temporal integration is performed strictly on the slow time scale $T$ with a 2nd-order Runge--Kutta method.  At each time step, the shape and angular frequency of the standing acoustic-wave mode are obtained by numerically solving the eigenvalue problem (\ref{wave:ode_g}) derived in \S~\ref{S2_1} using a Chebyshev spectral method and global QZ algorithm.  The amplitude of the acoustic mode $A$ either is taken to be a prescribed time-independent value or evolved according to the amplitude equation with an additional constant input power $P$, resulting in a term $P/\overline{E}_K$ on the right-hand side of (\ref{wave:amplitude_equation}).  In the first scenario, the forcing must be imagined to be tuned continuously to maintain the constancy of the wave amplitude, and the amplitude equation can then be utilised to self-consistently evaluate the power $P$ required. In the second scenario, in which the input power $P$ is held constant, (\ref{wave:amplitude_equation}) must be co-evolved numerically with the streaming equations.
%  a control loop would also be required since the frequency of the acoustic field evolves in time.

Depending upon the parameter regime, numerical integration of (\ref{SF:momentum})--(\ref{SF:state}) can result in a grid-scale instability manifesting as high-wavenumber variability in the $x$ direction.  To remedy this difficulty, we regularize the streaming dynamics simply by including the next-order terms in $\epsilon$ that, while formally small, introduce a viscous term of the form $\partial_{xx} \bar{u}_1$ in (\ref{SF:momentum}) and eliminate the numerical instability.

\begin{figure}
%  \centerline{\includegraphics[scale=0.5]{A2p8}}
    \centerline{\includegraphics[width=0.9\linewidth]{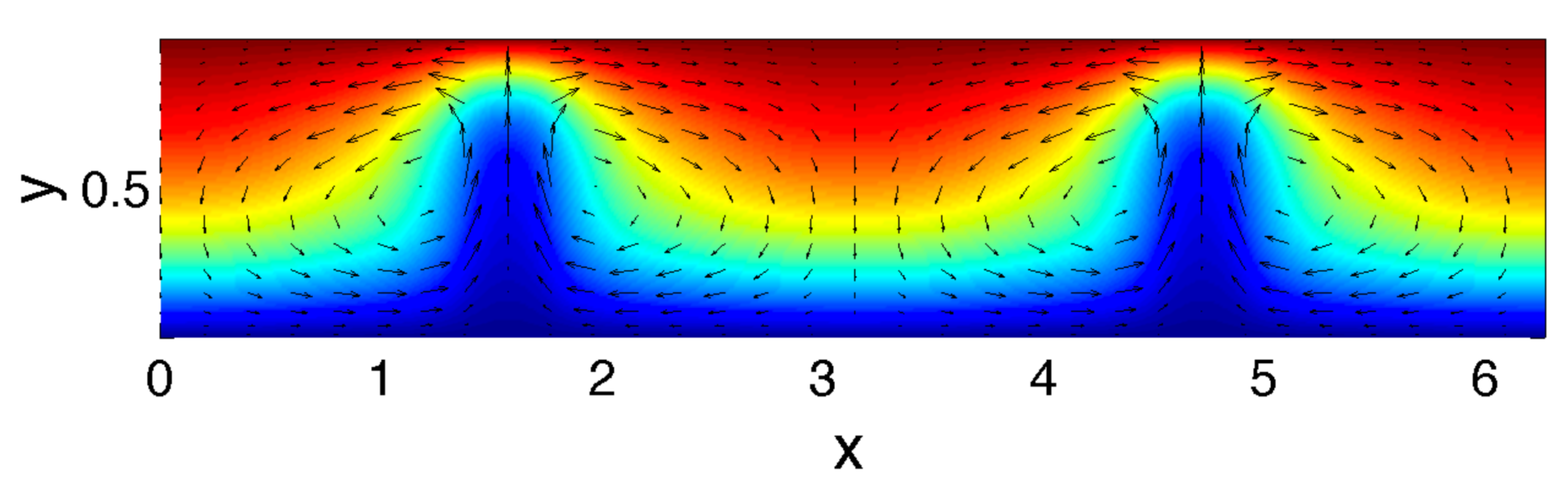}}
%        \centerline{\includegraphics[width=0.9\linewidth]{Fixed_A_Streaming.pdf}}
  \caption{Steady-state streaming velocity (vector arrows) and total temperature (color) fields, ($\bar{u}_1$,$\bar{v}_1$) and $T_B+\bar{\Theta}_0$, respectively, resulting from forcing at fixed acoustic-wave amplitude $A=2.8$.  The other parameters are chosen as follows:  $\epsilon=10^{-2}$, $\Gamma = 1$, $\gamma = 1.4$, $h = 4$, $Re_s = 4$ and $Pe_s = 4$.  Observe the emergence of vertical jets.}
\label{fig:A2p8}
\end{figure}

Here, we describe the the numerical simulations performed at constant amplitude $A$. The other parameters are chosen as follows:
\begin{equation}
\epsilon = 10^{-2},~~~ \gamma = 1.4, ~~~ \Gamma = 1, ~~~ h = 4, ~~~ Re_s = 4, ~~~ Pe_s = 4.\label{eq:numerical_parameters}
\end{equation}
Although not depicted here, for small values of $A$ the resulting steady-state velocity field agrees quantitatively with the one computed in \S~\ref{S3}. As $A$ is increased, however, the cellular streamline pattern evident in figure~\ref{fig:mode1_mean_velocity} changes into one dominated by narrow jet-like flow structures; see figure~\ref{fig:A2p8}, which shows the steady-state fields obtained for $A=2.8$.  Unlike streaming at high $Re_s$ in the absence of background density stratification \citep{Vainshtein1995}, symmetry with respect to the mid-plane $y=1/2$ is broken.  Moreover, for this simulation, the angular frequency of the acoustic waves changes by more than $3\%$, providing quantitative evidence of the two-way coupling between the waves and the streaming flow.

%%%%%%%%%%%%
%This is another source of difference between the solution at high $Re_s$ obtained in the absence of stratification, that conserves a symmetry with respect to $y=0.5$, and the one evidenced here. We also note that for this run, the angular frequency of the acoustic waves evolves of more than $3\%$, which evidences the coupling between the streaming-flow and the acoustic waves.

\begin{figure}
  \centerline{\includegraphics[scale=.75]{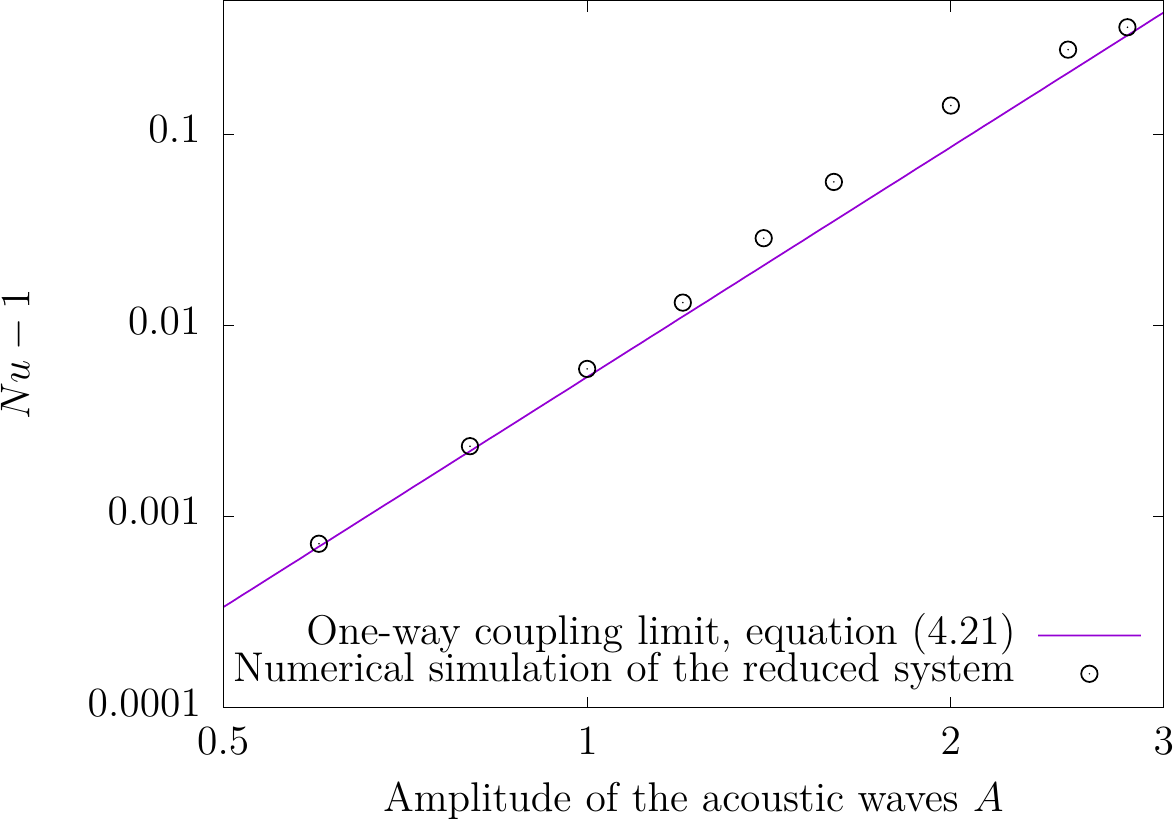}}
%    \centerline{\includegraphics[scale=.75]{Nu_vs_A}}
  \caption{Nusselt number $Nu$ versus forcing wave amplitude $A$ for baroclinic acoustic streaming in the small aspect-ratio limit without and with two-way wave/mean-flow coupling.
\label{fig:Nu_vs_A}}
\end{figure}

The heat-flux enhancement factors achieved by steady streaming states obtained for various values of $A$ (and computed using (\ref{eq:heat_flux_def})) are plotted in figure~\ref{fig:Nu_vs_A}. For small values of $A$, the results follow the theoretical prediction derived in \S~\ref{S3}, but ultimately deviate as the streaming flow develops localized jets.  Even in the regime of strong coupling, the rapid growth of the Nusselt number ($Nu-1 \propto A^4$) is qualitatively observed, strongly suggesting that significant increases in heat transport may be realized for sufficiently large values of $A$ (see \S~\ref{S5}).

%
%\\
%
%
%
\section{Discussion \label{S5}}

Both experiments and direct numerical simulations confirm that an imposed temperature difference or, more generally, any inhomogeneous background density field strongly affects both the pattern and intensity of acoustic streaming.  To correctly predict the streaming dynamics, it is thus necessary to properly account for these density gradients, as in the work of \cite{Chini2014} and in the present study.  The striking effect of the inhomogeneous density field on the streaming originates from the baroclinic generation of wave vorticity that occurs even in the absence of viscous torques.  This vorticity generation mechanism is readily inferred by taking the curl of the linearised Euler equations describing the leading-order acoustic wave dynamics,
\begin{equation}
\tilde{\mathbf{\nabla}} \times  \left( \tilde{\rho} \partial_{\tilde{t}} \tilde{\mathbf{u}} = - \tilde{\mathbf{\nabla}} \tilde{p} \right) \Rightarrow \partial_{\tilde{t}} \left(  \tilde{\mathbf{\nabla}} \times  \tilde{\mathbf{u}} \right) = \frac{( \tilde{\mathbf{\nabla}} \tilde{\rho}) \times (\tilde{\mathbf{\nabla}} \tilde{p})}{\tilde{\rho}^2},
\end{equation}
and from the instantaneous acoustic-wave velocity field shown in figure~\ref{fig:mode1_ac_field}.  As already emphasized in \cite{Chini2014}, this mechanism enhances the transfer of energy from the waves to the streaming flow; accordingly, this regime is appropriately termed \emph{baroclinic acoustic streaming}.

A quantitative understanding of baroclinic acoustic streaming is necessary for ascertaining the extent to which acoustics can be used to improve the transport and mixing of heat or of any dense or light solute chemical species. The separation in time scales between the period of the waves and the dynamics of the streaming flow, typically $\gtrsim10^3$, renders multiple scale analysis very attractive if not essential. The analysis reported here has been shown to accurately describe acoustic streaming in a thin channel across which a temperature difference is imposed: the explicit solution derived in the limit of one-way coupling quantitatively fits the data of a previous direct numerical simulation of the instantaneous governing equations (cf. \S~\ref{sec:comp}).  In the same limit, our analysis enables us to readily compute the heat flux enhancement engendered by the streaming and, in particular, to identify a very strong dependence of $Nu$ on the dimensionless channel height $h$:  $Nu-1\propto h^8$ (see (\ref{one_way:heat_flux})). This result clearly indicates that, to enhance the heat flux in a channel, the frequency of the acoustic wave should be tuned so that the aspect ratio $\delta$ is of order unity. 

The description of the $\delta = O(1)$ scaling regime also should be amenable to multiple scale analysis, an extension we are currently pursuing.  In this regime, baroclinic streaming continues to dominate viscous streaming, yet it is possible to more naturally account for viscous losses in the oscillatory boundary layers.  Indeed, whereas the \emph{streaming flow} induced by viscous dissipation in the Stokes boundary layers can be shown to be of higher order in the reduced system \citep{Chini2014}, an order of magnitude estimate of the dimensional time scale $\tilde{\tau}_{BL}$ over which the energy dissipation in the boundary layers damps the waves is given by
\begin{equation}
\tilde{\tau}_{BL} \sim \frac{(k_*^{-1}H_*) \times \rho_* (a_* \epsilon)^2 }{(k_*^{-1}\delta_{BL})\times \mu_* (a_* \epsilon/\delta_{BL})^2} \sim \frac{h \sqrt{Re_s}}{\omega_* \sqrt{\epsilon}},
\end{equation}
where $\delta_{BL} \sim \sqrt{\mu_* / (\rho_* a_* k_*)}$ is the thickness of the Stokes layers. Therefore, viscous dissipation in the oscillatory boundary layers takes place on a time scale that falls intermediate to the slow and the fast scales, and thus cannot be easily introduced within the framework of the small aspect-ratio analysis. Here, these viscous losses are implicitly presumed to be offset by the wave forcing mechanism.   (Note that in classical Rayleigh streaming, i.e. in a homogeneous medium, the dissipation time scale also lies between the wave period and the time required for steady streaming to be established.)  Practically, this results in a difference between the heat flux at the upper and lower walls, and this difference is the input power necessary to sustain the out-of equilibrium state \citep{Lin2008}. With the scaling $\delta = O(1)$, however, viscous dissipation in the Stokes layers occurs on the streaming time scale ($\tilde{\tau}_{BL} \omega_* \propto 1/ \epsilon$) and therefore can be readily incorporated into the analysis.  
%In classical Rayleigh streaming (i.e. in a homogeneous medium), the acoustic waves would be damped on a time scale intermediate to the wave period and the time required for steady streaming to be established.

The present study extends the investigation of \cite{Chini2014}, in which a similar multiple scale analysis is performed to predict acoustic streaming in a thin channel with cold top and bottom boundaries and an imposed volumetric heat source, in three significant ways.  First, we have derived an analytical solution for baroclinic acoustic streaming in the limit of one-way coupling and $\mathit{O}(1)$ Prandtl number that quantitatively accords with prior direct numerical simulations. Secondly, we performed numerical simulations of the reduced system that captures the feedback from the streaming to the acoustic wave field, i.e. two-way wave/mean-flow coupling, as manifested by the deviation of the resulting heat flux enhancement from that predicted using the analytical solution. In this stronger streaming regime, there is a transition from a rather smoothly-varying cellular streaming flow for which the wave-induced Reynolds stress divergence is balanced by the mean viscous force to a flow exhibiting jet-like structures and thin (streaming) boundary layers, akin to the transition from the Rayleigh to Stuart streaming regimes.  The detailed flow pattern, which exhibits only upward jets, differs from that realized at large $Re_s$ in the absence of stratification. Finally, perhaps the primary advance of the present study is that the dynamics of a multiple time-scale quasilinear wave/mean-flow system has been reduced to dynamical evolution strictly on the slow time scale: the spatial structure of the wave field is determined at each coarse time step from the solution of a 1D eigenvalue problem  (\S~\ref{S2_1}), while the modal amplitude evolves according to (\ref{wave:amplitude_equation}).

\begin{figure}
% \centerline{\includegraphics[scale=.75]{Q_A_vsT_new}}
  \centerline{\includegraphics[scale=.75]{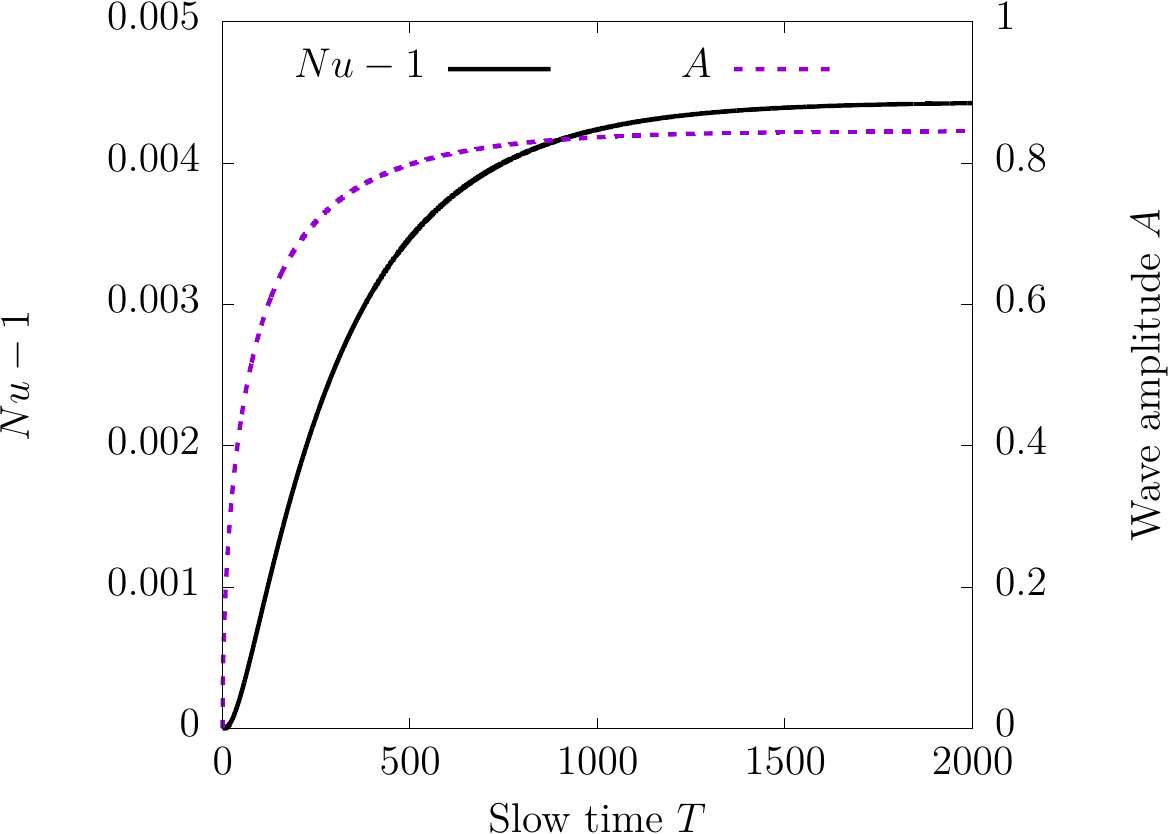}}
 \caption{Evolution of the Nusselt number $Nu$ and wave amplitude $A$ with slow time $T$ for baroclinic acoustic streaming with constant input power $P=10^{-3}$.  The remaining parameter values are specified in (\ref{eq:numerical_parameters}).}
\label{fig:NUAVST}
\end{figure}

This reduction yields immediate computational and theoretical advantages.  Indeed, by numerically integrating the streaming equations (\ref{SF:momentum})--(\ref{SF:state}) together with the wave amplitude equation (\ref{wave:amplitude_equation}) and eigenvalue problem (\ref{wave:ode_g}), the instantaneous dynamics need not be simulated using supercomputing resources  (e.g. as in \cite{Loh2002}) nor approximated by alternatively time-advancing the wave dynamics and the streaming flow \citep{Karlsen2017}.  Our algorithm thereby enables accurate and inexpensive numerical simulations over several thousand acoustic-wave periods to be performed in a regime where the waves and streaming flow are strongly coupled.  For example, in figure~\ref{fig:NUAVST}, we plot the evolution of the Nusselt number $Nu$ and the wave amplitude $A$ for baroclinic acoustic streaming with a small (dimensionless) constant input power $P=10^{-3}$.  Evidently, a steady state is not achieved until $T\approx 1000$, a time period corresponding to roughly $T/\epsilon = 10^5$ acoustic-wave cycles!  Even using modern computational capabilities, direct numerical simulation of the instantaneous Navier--Stokes equations for this scenario would be prohibitively expensive.  (For comparison, \cite{Lin2008} were able to run their case~1C simulations, discussed in \S~\ref{sec:comp}, for only a few hundered cycles.)  This computational challenge highlights the value of the asymptotically-reduced model, including the amplitude equation (\ref{wave:amplitude_equation}), which obviates the need for directly simulating the fast dynamics.  We note that the model of the quasi-biennial oscillation developed by \cite{Plumb1977} has certain similarities with the one derived here for baroclinic acoustic streaming:  in the former case, the internal gravity waves are explicitly determined using a WKBJ approximation, ultimately yielding a closed system for the mean (streaming) flow.  The analog of the amplitude equation (\ref{wave:amplitude_equation}), however, has not been derived, with the waves instead being held at constant amplitude.
%Finally, we describe baroclinic acoustic streaming when the input power is held constant.  Using the same parameter values as given in (\ref{eq:numerical_parameters}) and a dimensionless power input $P=0.001$, the time necessary to achieve a steady-state is $T\approx 1000$, see figure \ref{fig:NUAVST}. This time period corresponds to $T/\epsilon = 10^5$ acoustic cycles. Even using modern computational capabilities, direct numerical simulation of the instantaneous Navier--Stokes equations for this scenario would be prohibitively expensive.  (For comparison, \cite{Lin2008} were able to run their case~1C simulations, discussed in \S~\ref{sec:comp}, for only a few hundered cycles.)  This computational challenge highlights the value of the asymptotically-reduced model, including the amplitude equation (\ref{wave:amplitude_equation}), which obviates the need for directly simulating the fast temporal dynamics.
%subject to a constant power input. This new feature is made available by the derivation of the amplitude equation (\ref{wave:amplitude_equation}). This evidences the benefit of a reduced model in which the fast time-scale is eliminated from the dynamics: with the same parameters as (\ref{eq:numerical_parameters}) and a dimensionless power input $P=0.001$, the time necessary to achieve a steady-state is $T=  1000$, see figure \ref{fig:NUAVST}. This corresponds to $T/\epsilon = 100~000$ acoustic cycles. Even with high computational facilities, such an evolution could not have been obtained with direct numerical simulations of the full Navier-Stokes equations.
From a theoretical perspective, the derivation of the amplitude equation directly reveals the potential for a thermoacoustic-like instability in which an acoustic wave spontaneously grows in a stably stratified background, thereby increasing the heat flux.  It seems reasonable to assert that, without the amplitude equation, this instability would be difficult to anticipate and to mechanistically understand. 

%It may be of interest to attempt to adapt the multiscale analysis developed here to other QL systems.  The QL reduction increasingly is being invoked to study highly anisotropic (particularly sheared) turbulent flows, following a decomposition into a mean flow and eddies and a linearisation of the eddy dynamics. For instance, QL models have been used to describe the dynamics of zonal jets (see \cite{Marston2016} and references therein), the generation of magnetic fields in rotating astrophysical disks \citep{Squire2015}, the quasi-biennnial oscillation \citep{Plumb1977} and wall-bounded shear flow turbulence \citep{FARRELL}.
%strongly stratified turbulence \citep{Chini_inprep}. 
%The model of the quasi-biennial oscillation developed by \cite{Plumb1977} has certain similarities with the one derived here for baroclinic acoustic streaming:  in the former case, the internal gravity waves are described using a WKBJ approximation and explicitly determined, ultimately yielding a closed system for the mean (streaming) flow.  The analog of the amplitude equation (\ref{wave:amplitude_equation}), however, has not been derived, with the waves instead being held at constant amplitude.

%(according to the geophysical application).  Q:  What does this mean?

To conclude, we emphasize the importance of obtaining a quantitative understanding of the interaction of acoustic waves with a strongly stratified fluid. We are particularly interested in the potential heat-flux enhancement that can be achieved via baroclinic acoustic streaming in the absence of natural convection as a lightweight means of cooling electronics aboard spacecraft.  For this purpose, the analysis performed here should be extended to allow for channel aspect ratios of order unity.  For Earth-based applications, it also may be necessary to investigate the impact of buoyancy (gravitational) forces, which introduces into the  analysis another dimensionless parameter, the Richardson number $Ri=g_*/(k_*a_*^2)$, where $g_*$ is the acceleration of gravity.  Using the values $g_*=10~\mathrm{m}\cdot \mathrm{s}^{-1}$, $a_* = 333~\mathrm{m}\cdot \mathrm{s}^{-1}$ and $k_*=100~\mathrm{m}^{-1}$, we obtain $Ri = 10^{-6}$.  Accordingly, if $Ri$ were scaled in proportion to $\epsilon^{3/2}$, the mean-flow dynamics would be modified at leading-order in $\epsilon$; i.e. incorporation of gravity would, indeed, modify the results reported here, implying that quantitative experimental validation of the present study would require micro-gravity or microscale (microfluidic) environments.  Another important extension concerns the cooling of a hot object immersed in a fluid (e.g. again in the absence of natural convection) and subjected to acoustic-wave forcing.  As for the internal flow configuration studied here, a streaming flow is expected to develop and thereby enhance heat transfer.  To date, analysis of this fundamental problem, e.g. for a heated cylinder (first performed by \cite{Richardson1967} and \cite{ Davidson1973}; see \cite{Riley2001} for additional references), has been implicitly and, indeed, perhaps unknowingly restricted to small temperature anomalies.  Again, if the temperature differential between the object and the ambient fluid is sufficiently large, fundamental changes in the streaming dynamics may be anticipated with important ramifications for mixing rates and heat transfer that have yet to be properly explored.  
%The numerical or theoretical study of a jet-like boundary layer structure that develops at high streaming Reynolds number would help clarify this point. 
Above all, very few experiments of the baroclinic acoustic streaming regime have been performed: well-controlled experiments in micro-gravity or stably stratified systems would be invaluable for quantitative assessment of the theory.  Finally, a complementary perspective on acoustic streaming in stratified flows was recently offered by \cite{Beisner2015}, who provided evidence that acoustically-induced mixing can suppress combustion in micro-gravity environments. This phenomenon, too, relies on the strong coupling between a density stratified fluid and an acoustic wave.

\section*{Acknowledgements}

The authors are grateful to Keaton Burns for sharing his expertise on the Dedalus computing environment and for support from the 2017 Woods Hole Summer Program in Geophysical Fluid Dynamics (Woods Hole, MA), where much of this work was completed.

\appendix
\section{}\label{appA}
In this appendix, we report the dimensionless set of equations for the waves at second order. (The leading order equations are given in (\ref{waves:momentum_xy})--(\ref{waves:state}); also see \cite{Chini2014}.) In these equations, the linear operator acting on the $\mathit{O}(\epsilon^2)$ fluctuation fields is identified on the left-hand side and the inhomogeneous forcing arising from the leading-order terms is collected on the right-hand side. 
%These $\mathit{O}(\epsilon^2)$ fluctuation equations are given below.

\begin{enumerate}
\item The $x$-component of the momentum equation is
\begin{eqnarray}\nonumber
\omega_0 \bar{\rho}_0 \partial_\phi u_2' + \frac{1}{\gamma} \partial_x \pi_2' =- \omega_0 \left[ \rho_1 \partial_\phi u_1' - \overline{\rho_1' \partial_\phi u_1'} \right] - \omega_1 \bar{\rho}_0 \partial_\phi u_1' \\- \bar{\rho}_0 \left[ \partial_T u_1' + u_1 \partial_x u_1  - \overline{u_1 \partial_x u_1} + v_1 \partial_y u_1 - \overline{v_1 \partial_y u_1}  \right]  + \frac{1}{Re_s h^2} \partial_{yy} u_1'.\label{waves:second_order_vx}
\end{eqnarray}
\item The $y$-component of the momentum equation is
\begin{equation}
\partial_y \pi'_2 = -\gamma h^2 \bar{\rho}_0 \omega_0 \partial_\phi v_1'.\label{waves:second_order_vy}
\end{equation}
\item Conservation of mass requires
\begin{equation}
\omega_0 \partial_\phi \rho_2' + \partial_x (\bar{\rho}_0u_2') + \partial_y (\bar{\rho}_0v_2') = -\omega_1 \partial_\phi \rho_1' - \partial_T \rho_1' - \partial_x \left( \rho_1 u_1 - \overline{\rho_1 u_1} \right) - \partial_y \left(\rho_1 v_1 - \overline{\rho_1 v_1}\right).
\label{waves:second_order_conservation_mass}
\end{equation}
\item The internal energy balance is 
\begin{eqnarray}\nonumber
&&\omega_0 \partial_\phi \Theta_2' + u_2' \partial_x \bar{\Theta}_0 + v_2' \partial_y \left(\bar{\Theta}_0 + T_B \right) + (\gamma -1 ) \left[(T_B + \bar{\Theta}_0) (\partial_x u_2' + \partial_y v_2')\right] =\\&& - \omega_1 \partial_\phi \Theta_1' - \partial_T \Theta_1'  - u_1 \partial_x \Theta_1 + \overline{u_1 \partial_x \Theta_1}  - v_1 \partial_y \Theta_1 + \overline{v_1 \partial_y \Theta_1} \label{waves:second_order_energy_balance} \\&&+
(1-\gamma) \left[\Theta_1 (\partial_x u_1 + \partial_y v_1 ) - \overline{\Theta_1 (\partial_x u_1 + \partial_y v_1 )} \right] + \frac{\gamma}{\bar{\rho}_0 Pe_sh^2} \left(\partial_{yy} \Theta_1'- \frac{\rho_1' \partial_{yy} \bar{\Theta}_0}{\bar{\rho}_0} \right).  \nonumber
\end{eqnarray}
\item The equation of state is
\begin{equation}
\pi_2' - \rho_2' (T_B + \bar{\Theta}_0)- \bar{\rho}_0 \Theta_2' = \rho_1 \Theta_1 - \overline{\rho_1 \Theta_1}. \label{waves:equation_state}
\end{equation}
\end{enumerate}
\section{}\label{appB}
We can reduce the set of equations given in Appendix~\ref{appA} to (\ref{wave_2:eq1})--(\ref{wave_2:eq2}). This procedure is strictly analogous to the corresponding manipulations performed at leading order. The imaginary parts of $\mathcal{F}$ and $\mathcal{G}$, respectively $\mathcal{F}_i$ and $\mathcal{G}_i$, are found to be
\begin{eqnarray}\label{waves:FI}
\mathcal{F}_i &=& \frac{\omega_0}{\gamma} \partial_x \mathcal{H}_r - \frac{\omega_0 A(T)}{Re_s h^2} \partial_{yy} \hat{u}_1 + \omega_0 \bar{\rho}_0  \frac{\mathrm{d}A}{\mathrm{d}T} \hat{u}_1\\&&+\,\;\omega_0 \bar{\rho}_0 A(T)\left( \frac{\delta \hat{u}_1}{\delta T} + \hat{u}_1 \partial_x \bar{u}_1 + \bar{u}_1 \partial_x \hat{u}_1 + \hat{v}_1 \partial_y \bar{u}_1 + \bar{v}_1 \partial_y \hat{u}_1 \right)\nonumber ,
\end{eqnarray}
and
\begin{equation}
\mathcal{G}_i = \left(\frac{\omega_0}{\gamma}\right)\partial_y\mathcal{H}_r, \label{waves:GI}
%\frac{\partial_y \mathcal{H}_r \omega_0}{\gamma}, 
\end{equation}
%
%where $\delta \hat{f}/ \delta T$ is the functional derivative of $\hat{f}$ with respect to any dependence it may have on the slow time $T$, and $\mathcal{H}_r$ is defined by
where $\delta f/\delta T = (\partial \bar{\rho}_0 /\partial T) (\delta f/\delta \bar{\rho}_0)$ is shorthand notation for a differentiation operation involving the functional derivative $\delta f/\delta \bar{\rho}_0$. (Strictly, $\delta f/\delta T$ should read $\mathrm{d}f/\mathrm{d}T$, but the $\delta$-notation is used to emphasize that functional differentiation is involved.)  $\mathcal{H}_r$ is defined by
\begin{eqnarray}\nonumber
\mathcal{H}_r = \frac{i}{\omega_0\bar{\rho}_0 } \frac{\mathrm{d}A}{\mathrm{d}T} \hat{\rho}_1  +\frac{i}{\omega_0\bar{\rho}_0 } A(T) \left[ \frac{\delta \hat{\rho}_1}{\delta T} + \partial_x (\hat{\rho}_1\bar{u}_1) + \partial_y (\hat{\rho}_1\bar{v}_1)    \right]\hspace{1.0in}{}
\\\nonumber +\frac{iA(T)\bar{\rho}_0}{\omega_0} \left(\frac{1}{A(T)}\frac{\mathrm{d}A}{\mathrm{d}T} \hat{\Theta}_1 +  \frac{\delta \hat{\Theta}_1}{\delta T}  + \bar{u}_1 \partial_x \hat{\Theta}_1  + \bar{v}_1 \partial_y \hat{\Theta}_1\nonumber \right)\hspace{0.95in}{}\\+
\frac{iA(T)\bar{\rho}_0}{\omega_0}(\gamma -1) \left[\hat{\Theta}_1 (\partial_x \bar{u}_1 + \partial_y \bar{v}_1 ) \right] - \frac{iA(T)}{\omega_0} \frac{  \gamma}{Pe_sh^2} \left( \partial_{yy} \hat{\Theta}_1 - \frac{\hat{\rho}_1}{\bar{\rho}_0} \partial_{yy} \bar{\Theta}_0 \right) .
\end{eqnarray}

\section{}\label{appC}

In this appendix, we derive the solvability condition for the system (\ref{wave_2:eq1})--(\ref{wave_2:eq2}). In the vector space $(\mathbb{R}^2 \rightarrow \mathbb{C})^2$, a vector
\begin{equation}
\mathbf{V} =  \left(
\begin{array}{c}
  \hat{u}  \\
  \hat{v}
\end{array}  \right)  .
\end{equation}
The linear operator
\begin{equation}
\mathcal{L}=\left(  \begin{array}{cc}
  \partial_{xx} + \omega_0^2 \bar{\rho}_0 & \partial_{xy} \\
  \partial_{xy} & \partial_{yy}
\end{array} \right),
\end{equation}
so the set of equations (\ref{wave_2:eq1})--(\ref{wave_2:eq2}) is of the form $\mathcal{L}\mathbf{V}  = \mathbf{F} $.  Given two vectors $\mathbf{V}_A$ and  $\mathbf{V}_B$, we define their scalar product  
%$(\cdot \vert \cdot )$ as
%
\begin{equation}
\left( \mathbf{V}_A  \vert \mathbf{V}_B \right) \equiv \int_0^{2\pi} \mathrm{d} x \int_0^1 \mathrm{d}y \left(\mathbf{V}_A^\mathrm{T} \cdot \mathbf{V}_B^* \right)= \int_0^{2\pi}\mathrm{d}x \int_0^1 \mathrm{d}y \left( \hat{u}_A\hat{u}_B^* + \hat{v}_A \hat{v}_B^* \right),
\end{equation}
where $\hat{f}^*$ stands for the complex conjugate of $\hat{f}$. Given the $2\pi$-periodicity requirement in $x$ and the kinematic boundary conditions in $y$, the operator $\mathcal{L}$ is self-adjoint ($\mathcal{L}^\dagger= \mathcal{L}$); i.e.
\begin{equation}
\left( \mathcal{L} \mathbf{V}_A \vert \mathbf{V}_B \right) = \left(  \mathbf{V}_A \vert \mathcal{L} \mathbf{V}_B \right) .
\end{equation}
Therefore, the vectors in the kernel of $\mathcal{L}$ consist of the $\mathit{O}(\epsilon)$ acoustic-wave modes already described.  Since one of these modes $\mathbf{V}_1=(\hat{u}_1, \hat{v}_1)$, then
\begin{equation}
\left( \mathbf{F} \vert \mathbf{V}_1 \right) = 0.\label{raw_solvability} 
\end{equation}
The imaginary part of (\ref{raw_solvability}) requires
\begin{equation}
\int \mathrm{d}x \mathrm{d}y \left( \mathcal{F}_i \hat{u}_1 + \mathcal{G}_i \hat{v}_1 \right) = 0, \label{solvability}
\end{equation}
recalling that both $\hat{u}_1$ and $\hat{v}_1$ are real-valued (so $\hat{u}_1 = \hat{u}_1^*$ and  $\hat{v}_1 = \hat{v}_1^*$).

\section{}\label{appD}

Evaluation of the solvability condition (\ref{solvability}) requires functional forms for $\hat{u}_1$ and $\hat{v}_1$, as detailed in  (\ref{wave:uhat})--(\ref{wave:ode_g}), and $\mathcal{F}_i$ and $\mathcal{G}_i$, whose explicit expressions are reported in (\ref{waves:FI})--(\ref{waves:GI}). Obtaining (\ref{wave:amplitude_equation}) from (\ref{solvability}) is a lengthy but straightforward calculation, except for the treatment of the functional derivatives.  Certain functional derivatives can be simplified by exploiting the normalization condition; we illustrate this technical point through the computation of two integrals necessary to derive (\ref{wave:amplitude_equation}),
\begin{equation}
I_1= 2 \int_0^{2\pi} \mathrm{d}x g \frac{\delta g}{\delta T},
\end{equation}
and
\begin{equation}
I_2= \int_0^{2\pi} \mathrm{d}x \int_0^1 \mathrm{d}y \hat{u}_1 \bar{\rho}_0 \frac{\delta \hat{u}_1}{\delta T}.
\end{equation}

For a real-valued functional $J$ of a function $j$, the functional derivative $\delta J / \delta j$ describes how $J(j)$ evolves when $j \rightarrow j + w$ with $\vert w \vert \rightarrow 0$:
\begin{equation}
\frac{\delta J}{\delta j} (w) = \lim_{\varepsilon \rightarrow 0} \left( \frac{J(j + \varepsilon w) - J(j)}{\varepsilon} \right).
\end{equation}
In the present work, the functions we consider ($\hat{u}_1$, $\hat{v}_1$, etc.) have a functional dependence on $\bar{\rho}_0$ and also depend on the real parameters $x$ and $y$.  Specifically, we are interested in how the quantity $\hat{f}(x,y; \bar{\rho}_0)$ evolves at a fixed position $x$ and $y$, while $\bar{\rho}_0$ evolves slowly in time. Formally, we therefore should define a functional $\hat{f}_{x,y}$ such that
\begin{equation}
  \hat{f}_{x,y} ~\left\{
    \begin{array}{ll}
      ([0,2\pi ] \times [0,1] \rightarrow \mathbb{R}) \rightarrow \mathbb{C}\\
      \bar{\rho}_0 \rightarrow \hat{f}(x,y;\bar{\rho}_0)
  \end{array} \right.
\end{equation} 
and the shorthand $\delta \hat{f} /\delta T$ should therefore be understood as
\begin{equation}
\frac{\delta \hat{f}}{\delta T} \rightarrow \frac{\delta \hat{f}_{x,y}}{\delta \bar{\rho}_0}(\partial_T \bar{\rho}_0 ).
\end{equation}

The first integral $I_1$ is thus given by
\begin{equation}
I_1 = \int_0^{2\pi} \mathrm{d}x  \frac{\delta g_x^2}{\delta \bar{\rho}_0}(\partial_T \bar{\rho}_0 ) = \lim_{\varepsilon \rightarrow 0} \left(\frac{\int_0^{2\pi} \mathrm{d}x g_x^2(\bar{\rho}_0 + \varepsilon \partial_T \bar{\rho}_0) - \int_0^{2\pi} \mathrm{d}x g_x^2(\bar{\rho}_0)}{\varepsilon} \right).\label{I1simplified}
\end{equation}
According to the normalization condition (\ref{wave:normalization}), each of the integrals on the right-hand side of (\ref{I1simplified}) is equal to unity so that $I_1 = 0$.

Similarly, the second integral is expressed as a function of $g$ with (\ref{wave:uhat}),
\begin{eqnarray}
I_2 &&= \frac{1}{\omega_0^2} \int_0^{2\pi} \mathrm{d}x \int_0^1 \mathrm{d}y g' \frac{\delta}{\delta T} \left( \frac{g'}{\omega_0^2 \bar{\rho}_0} \right)\\
&&= \frac{1}{\omega_0^2} \int_0^{2\pi} \mathrm{d}x \int_0^1 \mathrm{d}y g'  \left[ \frac{1}{\omega_0^2 \bar{\rho}_0} \frac{\delta g'}{\delta T} - \frac{2 g' \partial_T \omega_0}{\bar{\rho}_0 \omega_0^3} - \frac{g' \partial_T \bar{\rho}_0}{\omega_0^2 \bar{\rho}_0^2} \right].\label{I2simplified}
\end{eqnarray}
The various contributions to (\ref{I2simplified}) can be simplified using the function $\alpha$ introduced in (\ref{alpha}) since only $\bar{\rho}_0$ depends on $y$:
\begin{equation}\label{appendix_i2}
I_2 =  \frac{1}{\omega_0^4} \int_0^{2\pi} \mathrm{d}x g' \alpha \frac{\delta g'}{\delta T} - \frac{2\partial_T \omega_0}{\omega_0^5} \int_0^{2\pi} \mathrm{d}x \alpha g'^2  +\frac{1}{\omega_0^4} \int_0^{2\pi} \mathrm{d}x  g'^2 \partial_T \alpha  .
\end{equation}
The normalization condition (\ref{wave:normalization}) with the constitutive relation for $g$ (\ref{wave:ode_g}) yields
\begin{equation}
\int_0^{2\pi}\mathrm{d}x \alpha g'^2  = \left[ \alpha g g' \right]_0^{2\pi} + \int_0^{2\pi} \mathrm{d}x \omega_0^2 g^2 = \omega_0^2.
\end{equation}
This result can be used to compute the first integral in (\ref{appendix_i2}), 
\begin{eqnarray}
\int_0^{2\pi} \mathrm{d}x \alpha &&g' \frac{\delta g'}{\delta T} = \frac{1}{2} \int_0^{2\pi} \mathrm{d}x \left( \frac{\delta (\alpha g'^2)}{\delta T} - g'^2 \frac{\delta \alpha}{\delta T} \right)\\
&&= \frac{1}{2} \left[ \lim_{\varepsilon \rightarrow 0} \left( \frac{\int_0^{2\pi} \mathrm{d}x (\alpha g'^2)(\bar{\rho}_0 + \varepsilon \partial_T \bar{\rho}_0)   -\int_0^{2\pi} \mathrm{d}x (\alpha g'^2)(\bar{\rho}_0)      }{\varepsilon}   \right) -  \int_0^{2\pi} \mathrm{d}x g'^2 \partial_T \alpha       \right] \nonumber \\
&&= \frac{1}{2} \left[ 2 \omega_0 \partial_T \omega_0 -  \int_0^{2\pi} \mathrm{d}x g'^2 \partial_T \alpha       \right]  =\omega_0 \partial_T \omega_0 - \frac{1}{2} \int_0^{2\pi} \mathrm{d}x g'^2 \partial_T \alpha.      \nonumber
\end{eqnarray}
Finally, we obtain
\begin{equation}
I_2 = - \frac{\partial_T \omega_0}{\omega_0^3} +  \frac{1}{2\omega_0^4} \int_0^{2\pi} \mathrm{d}x g'^2 \partial_T \alpha .
\end{equation}

\bibliographystyle{jfm}

\end{document}